\documentclass[lettersize,journal]{IEEEtran}

\usepackage[utf8]{inputenc}
\usepackage[T1]{fontenc}
\usepackage{amsmath,amssymb,amsfonts}
\usepackage{graphicx} 
\usepackage{xcolor}
\usepackage{algorithm}
\usepackage{algorithmic}
\usepackage{array}
\usepackage{textcomp}
\usepackage{stfloats}
\usepackage{url}
\usepackage{verbatim}
\usepackage{cite}
\usepackage{authblk}
\usepackage{setspace}
\usepackage{booktabs}
\usepackage{caption}
\usepackage{subcaption} 
\usepackage{float} 
\usepackage[colorlinks=true,citecolor=blue,linkcolor=blue,urlcolor=blue]{hyperref} 
\usepackage[nohyperlinks]{acronym}
\usepackage{siunitx}
\usepackage{tikz}
\usetikzlibrary{matrix,shapes.geometric} 

\begin{document}

\title{Multi-Segment Photonic Power Converters for Energy Harvesting and High-Speed Optical Wireless Communication}

\author{Othman~Younus,~\IEEEmembership{Member,~IEEE,}
        Behnaz~Majlesein,
        Richard~Nacke,
        Isaac~N.~O.~Osahon,
        Carmine~Pellegrino,
        Sina~Babadi,
        Iman~Tavakkolnia,
        Henning~Helmers,~\IEEEmembership{Member,~IEEE,}
        and~Harald~Haas,~\IEEEmembership{Fellow,~IEEE}%
\thanks{Othman Younus, Behnaz Majlesein, Isaac N. O. Osahon, Sina Babadi, Iman Tavakkolnia, and Harald Haas are with the LiFi Research and Development Center (LRDC), Electrical Division, Department of Engineering, University of Cambridge, Cambridge, CB3 0FA, United Kingdom.}%
\thanks{Richard Nacke, Carmine Pellegrino, and Henning Helmers are with the Fraunhofer Institute for Solar Energy Systems ISE, Heidenhofstraße 2, 79110 Freiburg, Germany.}%
\thanks{Corresponding author: Othman Younus (e-mail: oiyy2@cam.ac.uk).}
\thanks{Manuscript received October 25, 2025; revised August 26, 2025.}}

\maketitle
\setcounter{page}{1} 

\markboth{Journal of \LaTeX\ Class Files,~Vol.~1, No.~1, August~2025}%
{Younus \MakeLowercase{\textit{et al.}}: Multi-Segment Photonic Power Converters}


\begin{abstract}

The demand for energy-efficient high-speed wireless communication, coupled with the rapid rise of IoT devices, requires systems that integrate power harvesting with optical data reception to eliminate the need for charging or battery replacements. Recent advances have explored the use of solar cells as optical receivers for high-speed data detection alongside power harvesting. \acs{GaAs}-based \acp{PPC} provide six times greater electron mobility than silicon- or cadmium telluride-based cells, enabling faster data detection and improved power efficiency. However, their bandwidth is constrained by junction capacitance, which increases with active area, creating a trade-off between power output and data rate. To address this, we propose and test multi-segment \acs{GaAs}-based \Acp{PPC} that serve as both energy harvesters and data detectors. By segmenting the active area into 2, 4, or 6 subcells, forming circular areas with diameters of 1, 1.5, or 2.08~mm, we reduce capacitance and boost bandwidth while preserving light collection. Fabricated on a semi-insulating \ac{GaAs} substrate with etched trenches for electrical isolation, the series-connected subcells optimize absorption and minimize parasitic effects. The \Acp{PPC} were used for an eye-safe 1.5~m optical wireless link, employing \ac{OFDM} with adaptive bit and power loading. The system achieved a world record data rate of 3.8~Gbps, which is four times higher than prior works. The system converts 39.7\% of optical power from a beam of 2.3~mW, although the segmentation increases the sensitivity of the alignment. These findings provide new solutions for off-grid  backhaul for future communication networks, such as 6th generation (6G) cellular.

\end{abstract}


\section{Introduction}
The growing demand for high-speed, low-latency and energy-efficient connectivity continues to drive the evolution of global telecommunications\cite{ref1}. This growth is driven by the rapid expansion of \ac{IoT} devices, with projections indicating an increase from 19.8 billion connected worldwide in 2025 to 40.6 billion in 2034 \cite{ref5}. Powering these devices poses substantial challenges, as batteries have finite lifespans, requiring replacements that cause operational downtime, and deployments in remote or resource-constrained environments introduce further complexities.
Fiber-optic infrastructure remains the backbone of modern networks, yet its high deployment cost and susceptibility to geographical and logistical constraints limit scalability. \Ac{OWC}, including \ac{FSO}, has emerged as a compelling alternative for both terrestrial and space-based backhaul. Its high bandwidth, rapid deployment, and resilience to electromagnetic interference position it as a key enabler for network extension in hard-to-reach environments\cite{ref3}. In addition to data transmission, the \ac{OWC} links have the potential for integrated power delivery, positioning them as a dual-purpose platform for next-generation networks.

\Ac{SWIPT} enhances \ac{RF} offers a unified framework for data and energy transmission, but its performance depends heavily on the choice of carrier. \ac{RF}-based implementations are limited by low power density, rapid attenuation, and poor energy conversion efficiency at low signal levels \cite{wang2025novel}. In contrast, lightwave-based systems, collectively referred to as \ac{RF} systems, 
utilize optical carriers to enable both data transmission and power delivery, and have emerged as a promising alternative to conventional \ac{RF}-based solutions\cite{ref6, ref7}. \ac{RF} systems offer significantly higher directional power transfer efficiency and broader bandwidth, making them well-suited for compact, energy-constrained devices in dense or remote deployments. Key advantages include (i) extended range due to low divergence and reduced attenuation compared to \ac{RF} systems; (ii)  access to license-free, and high bandwidth optical channel; (iii) immunity to \ac{EM} interference \cite{ref1}; (iv) reduced net power consumption; (v) enhanced physical-layer security from the confined propagation of light beams; and (vi) compatibility with \ac{RF}-restricted environments \cite{Richard}.

\begin{figure}[!]
    \centering
\begin{tikzpicture}
    \node[draw, line width=1pt, rectangle, inner sep=2pt, rounded corners=5pt] (figbox) { 
        \newlength{\mycolumnwidth}
        \setlength{\mycolumnwidth}{0.32\textwidth} 
    \includegraphics[width=0.92\columnwidth, trim=0cm 6cm 0cm 6cm, clip]{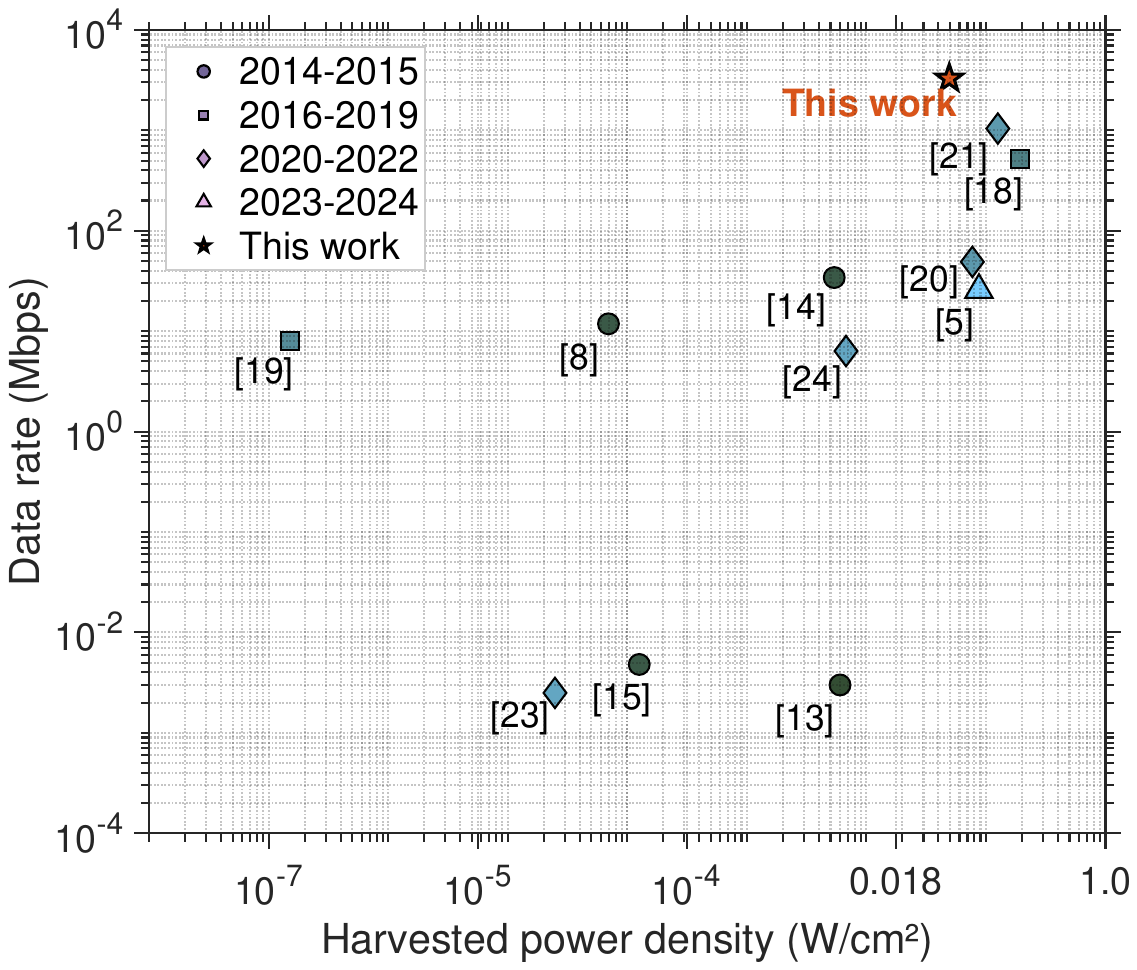}
    };
\end{tikzpicture} 

    \caption{Comparison of state-of-the-art studies on \ac{PV} cell-based receivers with this works result using a 4-segment \ac{GaAs}-based cell.}
    \label{fig:pv_cell_rx}
\end{figure}

The initial demonstration of \ac{SLIPT} systems utilized cost-effective, widely available \acf{Si}-based \ac{PV} cells for data reception \cite{ref8, ref9, ref10, Walker, ref12}. In \cite{ref8, ref12}, a short-distance communication link (up to one meter) was established using white \acp{LED} and multi-crystalline \ac{Si} \ac{PV} with \ac{OFDM} modulation. The system, constrained by a bandwidth of only a few MHz, demonstrated data rates of 7~Mbps and 12~Mbps at \ac{BER} thresholds of $6.6\times10^{-4}$ and $1.6\times10^{-3}$, respectively. Figure \ref{fig:pv_cell_rx} summarizes key studies on \ac{SLIPT} systems across various operational scenarios, highlighting the trade-off between the data rate and the harvested power density \cite{ref3, ref6, ref13, ref8, ref15, ref16, ref17, ref18, ref19, ref20, ref21, ref22, ref23, ref24, ref25}. The practicality of \ac{SLIPT}-enabled communication was demonstrated in outdoor conditions, such as the system deployed in the Orkney Islands, Scotland \cite{ref20}, where a real-time data throughput of 8 Mbps was achieved. This was made possible by the minimal alignment requirements enabled by the large active area of the \ac{Si} \ac{PV} receiver. However, the inherent bandwidth limitations of \ac{Si} \ac{PV} cells are attributed to their low carrier mobility \cite{ref22}.

III-V semiconductors offer distinctive optoelectronic properties and tunable bandgaps, making them highly suitable for \acp{PPC} compared to conventional materials like \ac{Si} \cite{ALGORA2022340}. Among III-V materials, \ac{GaAs} is particularly advantageous for light in the 800-870 nm range due to its direct bandgap of 1.42~eV, which yields a high optical absorption coefficient. This enables the design of thinner photovoltaic cells that efficiently absorb monochromatic laser light \cite{Walker}. In addition, lattice-matched growth on GaAs substrates enables high material quality, which along with the direct bandgap facilitates radiatively dominated recombination, reducing non-radiative losses and thereby enhancing conversion efficiency. Besides, \ac{GaAs} exhibits carrier mobility approximately six times higher than that of \ac{Si}, which supports faster frequency response for data transfer applications and reduces sheet resistance, mitigating secondary losses \cite{balaghi2021high}. The inherent tunability of the bandgap in III-V compounds also enables precise matching to the specific photon energy of incident radiation, thus reducing thermalization and transmission losses $(\Delta E_{\text{therm}} = E_{\text{photon}} - E_g)$. This tunability further improves overall conversion efficiency by overcoming the intrinsic limitations of spectrally mismatched absorbers by improved photon utilization \cite{Walker}.\\
These material advantages translate directly into device-level performance. \acp{PPC} based on III-V semiconductors, particularly \ac{GaAs}, have demonstrated significantly higher conversion efficiencies than typical solar cells. Unlike conventional photovoltaic devices optimized for the full solar spectrum, \acp{PPC} are specifically designed for monochromatic light sources, such as lasers, allowing for tailored, high-efficiency energy conversion \cite{ref32}. For instance, GaAs-based \acp{PPC} have achieved power conversion efficiencies (PCE) as high as 68.9\% under laser illumination at 858~nm \cite{ref27}. 

While silicon-based \acp{PPC} have demonstrated efficiencies of 9.9\% and 24\% under illumination at wavelengths of 980~nm and 1070~nm, respectively, their performance remains inferior to that of devices based on GaAs and other III-V compounds \cite{allwood2011comparison, Percentage15}. 

Overall, III-V-based \acp{PPC} offer faster response times and higher energy harvesting efficiency, positioning them as promising candidates for \ac{SLIPT} systems. 

The high performance of \ac{GaAs} enables \acp{PPC} to operate under high optical power densities, allowing miniaturization without compromising the total harvested power. This supports the use of small \acp{PPC} elements for high-speed operation, through reduced capacitance and faster response, while maintaining overall power output through the integration of the array. Notably, the size of individual \acp{PPC} should not be conflated with the total receiver area, which can comprise many such elements. The key trade-off lies in optimizing the dimensions of the device for bandwidth without sacrificing power harvesting \cite{ref22}.

Previously, a record data rate of 1.041~Gbps was achieved over a 2-meter distance using infrared laser transmitters, while also attaining a~41.7\% \ac{PCE} at an incident laser power density of 0.3~W/cm$^2$\cite{ref22}. In this work, we introduce multi-segment device architectures for energy harvesting and data reception in \acp{MIM} for \ac{SLIPT} systems. These devices divide a single \ac{PPC} cell into multiple electrically connected subcell segments in series. Hence, it offers an increased output voltage and reduces the capacitance of the device due to the reciprocal sum law of series-connected segments, enabling higher bandwidth for data reception. The experimental system demonstrated in this work employs an \ac{GaAs}-based \ac{PPC} with multiple monolithically interconnected subcells in series on the chip. This architecture yields lower output currents than unsegmented cells due to the smaller area of each segment and typically features an increased series resistance from lateral conduction required for interconnections but provides higher voltage and lower capacitance to improve \ac{SLIPT} performance.

To the best of our knowledge, this is the first demonstration of a multi-segment \ac{GaAs} \ac{PPC}-based receiver achieving a maximum data rate of 3.8~Gbps. In addition to high-speed reception, the system also converts up to 39.7\% of the received optical power into usable electrical energy. This advancement sets the foundation for future energy-autonomous optical wireless communication systems capable of sustained operation in both indoor and outdoor environments while reducing maintenance overhead and advancing the sustainability of next-generation wireless networks.

\section{Proposed System}
\label{sec:methods}

\subsection{Sample description}
\begin{figure*}[h!]
    \centering
    \begin{tikzpicture}
    \node[draw, line width=1pt, rectangle, inner sep=2pt, rounded corners=5pt] (figbox) {
        \begin{minipage}{\textwidth}
    \begin{subfigure}[b]{0.8\textwidth}
        \centering
        \includegraphics[width=\textwidth, trim=-15cm 13cm 0cm 27cm, clip]{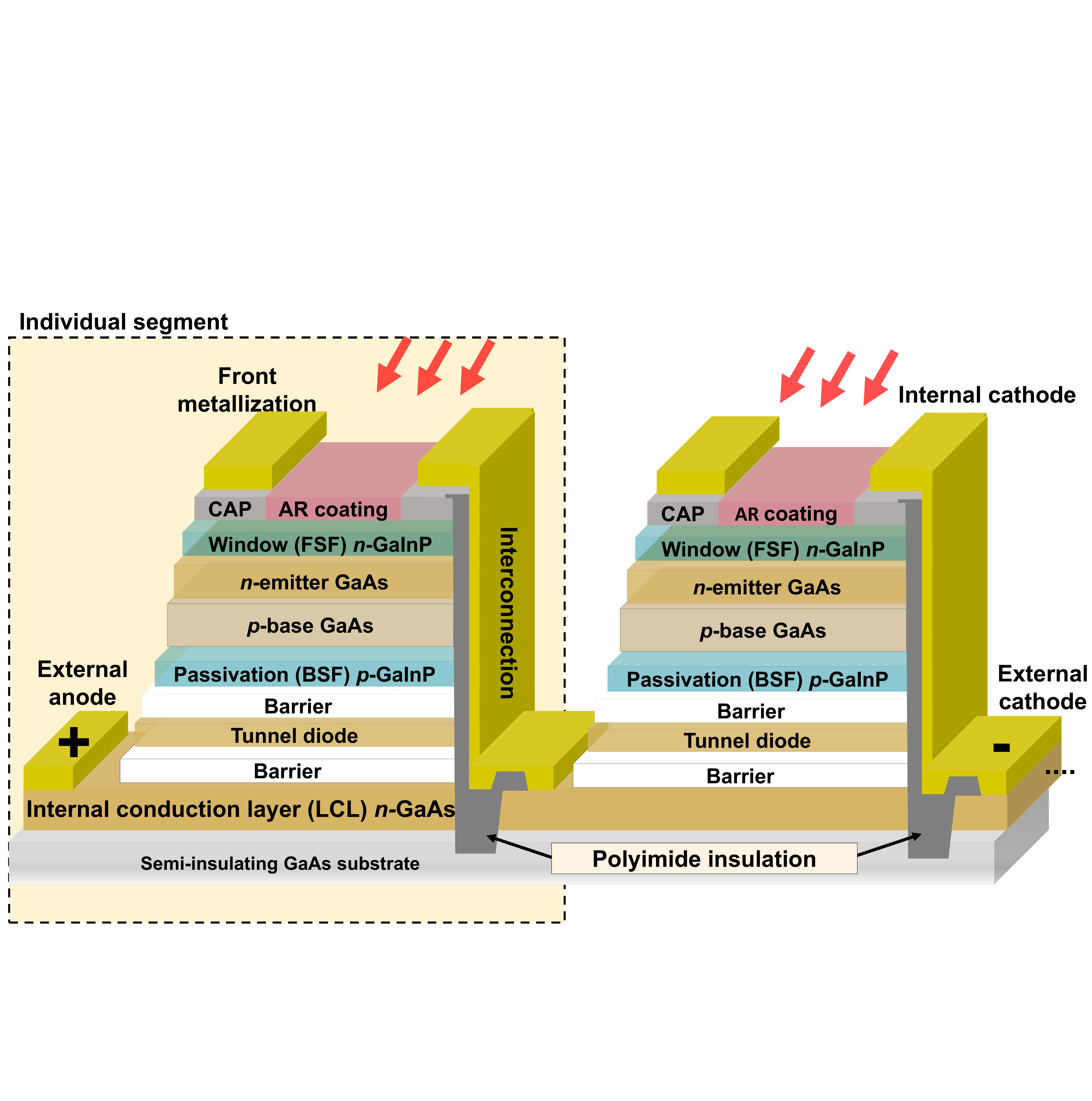}
        \caption{}
        \label{fig:GaAs_receiver_b}
    \end{subfigure}

    \vspace{1em} 

    \begin{subfigure}[b]{0.5\textwidth}
        \centering
         \includegraphics[width=\columnwidth, height=13cm, keepaspectratio, trim=1cm 7cm 2cm 8cm, clip]{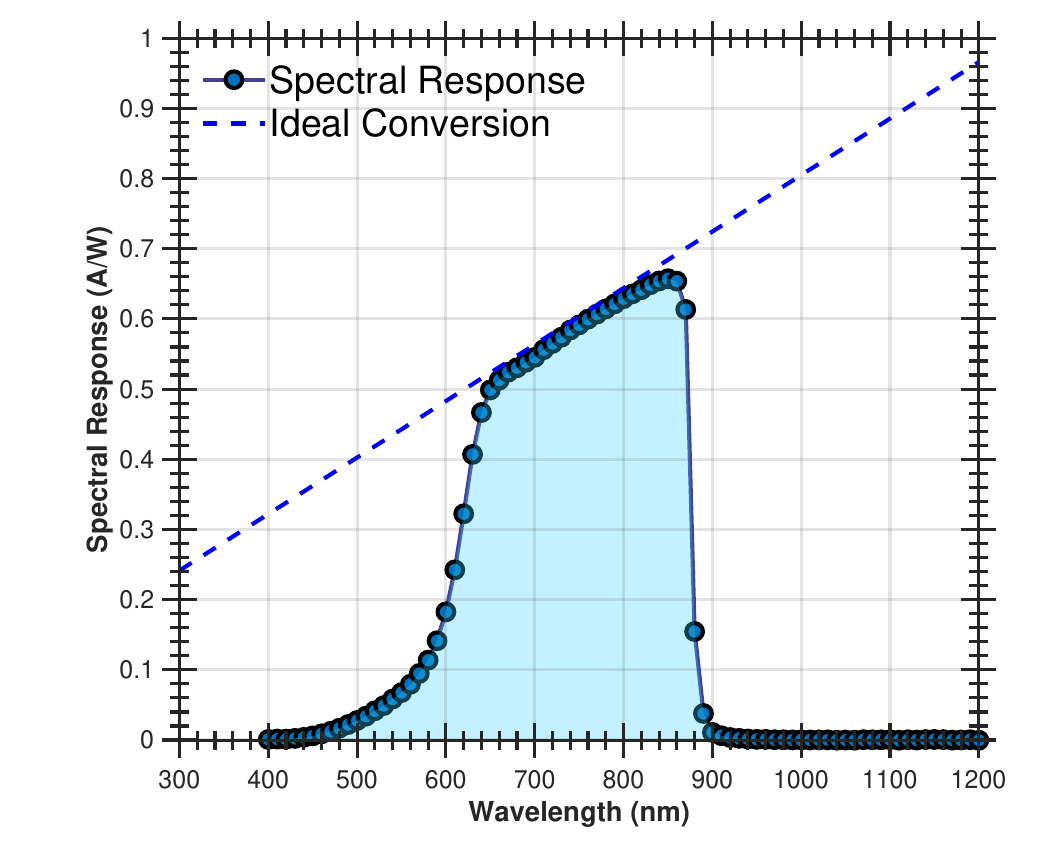}
        \caption{} 
        \label{fig:GaAs_receiver_c}
    \end{subfigure}
    \begin{subfigure}[b]{0.48\textwidth}
        \centering
     \includegraphics[width=\columnwidth, height=13cm, keepaspectratio, trim=0cm 5cm 0cm 6cm, clip]{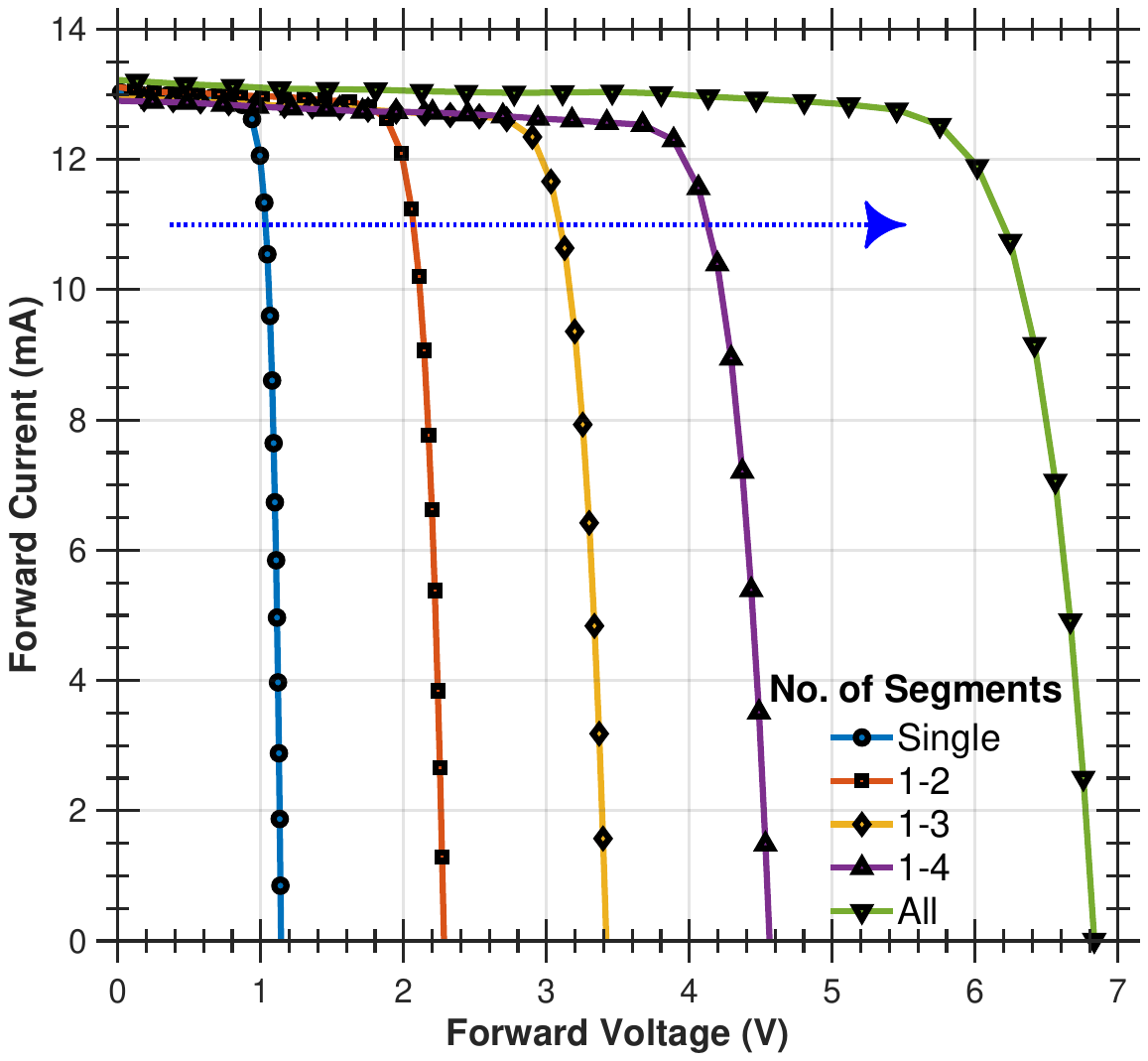}        
        \caption{} 
        \label{fig:GaAs_receiver_d}
    \end{subfigure}
        \end{minipage}
    }; 
    \end{tikzpicture} 

\caption{Multi-segment GaAs photovoltaic receiver structures: 
    \textbf{a,} Schematic of the epitaxial structure and interconnection scheme.
    \textbf{b,} Measured spectral response of a single segment.
    \textbf{c,} Example current-voltage curves of single- and only partially interconnected segments as well as of an entire 6-segment device measured under monochromatic 809-nm illumination at uniform and similar irradiance.}

    \label{fig:GaAs_receiver_structures} 
\end{figure*}

The experimental setup investigates the use of a GaAs-based multi-segment \ac{PPC} as \ac{Rx} with 2, 4, and 6 segments in three different sizes, with circular active areas with diameters of 1 (S), 1.5 (M), and 2.08~mm (L), fabricated at Fraunhofer ISE. The 2, 4, or 6 individual photovoltaic subcell segments per chip are electrically separated by isolation trenches into semi-insulating \ac{GaAs} substrate (\ac{MIM}). Outside the circular active area, partial areas are etched down to expose a lateral conduction layer which is implemented underneath the active pn-junction. By metal bridges across polyimide-filled trenches, adjacent circular sectors are interconnected in series in the so-called “pizza-configuration”. A colored micrograph of example devices is shown in Fig.~\ref{fig:Samplec}. Green and yellow areas denote the positive and negative contact areas of the rear and front, respectively. The chips were attached to a planar submount
and contacted via wire bonding to the two larger contact pads on one side of the chip. Phot-generated carriers are collected by front metal grid lines that transport electrons to the metal busbars at the rim. At the rear of the junction a tunnel diode is used to flip polarity, to allow lateral transport from the center to the rim underneath the pn-junction in a 5~$\mu$m thick n-type GaAs lateral conduction layer. A schematic layer structure is shown in Figure \ref{fig:GaAs_receiver_b}.

Each segment consists of the same semiconductor structure based on a GaAs pn-junction (3650~nm absorber). The front and back surfaces are passivated using n- and p-type \ac{GaInP} layers, respectively. The front surface field GaInP is 400~nm thick to support lateral transport to the grid lines, but is transparent to the monochromatic target wavelength of 850~nm. The epitaxial structure was grown on a 4-inch semi-insulating \ac{GaAs} substrate via \ac{MOVPE}. Post-epitaxial processing involves photolithographic micropatterning, selective wet etching, and polyimide passivation to isolate device mesas. The metallization of the anode and cathode front side is achieved by thermal evaporation, followed by the deposition of a double-layer \ac{ARC} to minimize optical losses. More details on this sample are available in \cite{ref38}, together with the fabrication procedure \cite{ref32}. The multi-segment approach allows for harvesting power from the series-connected photovoltaic cell device at elevated voltage, which is beneficial for downstream electronics and load matching. At the same time, the integrated series connection reduces the device capacitance due to the reciprocal sum law and thereby overcomes the signal-capacitance trade-off. The key metrics of the fabricated devices are assessed in terms of the spectral response, the illuminated \ac{I-V} characteristic, dark \ac{I-V} analysis, and sheet resistance extraction, which was reported in \cite{ref32}.

The device spectral response, shown in Figure \ref{fig:GaAs_receiver_c}, peaks near 850~nm, demonstrating its effectiveness as a \ac{PPC}. The response at 850~nm is ideal for converting light from common laser or LED sources at this wavelength into electrical power. The sharp drop-off at 870~nm results from the absorber being a GaAs semiconductor crystal with a bandgap of 1.42 eV. The  deviation from the ideal 100\% quantum efficiency (dashed line) originates from small optical losses (shading from metal grid, non-ideal anti-reflection coating). The absence of spectral dependencies, i.e. the following of the ideal curve, in the relevant range indicates good material quality and sufficient minority carrier diffusion length in the absorber material. In addition, \ref{fig:GaAs_receiver_d}  illustrates  the \ac{I-V} curves for single segments, and its accumulation for the interconnected 6-segment device, measured under uniform 809-nm illumination with similar but not calibrated optical irradiance. 

\subsection{Experimental setup}
\label{sec:Experimental setup}

The overall system design is presented in Fig \ref{fig:experimental_setup_photo}, in which a \ac{DC} power supply (Keysight, E36313A) is used to provide the required bias level for the transmitter (i.e., a voltage of 1.78~V and a current of 6~mA), which ensures the \ac{VCSEL} is operated in its linear range. This \ac{DC} power supply is additionally used for \ac{DC} measurements of the receiver \ac{I-V} curve. An \ac{AWG} (Keysight, M8195A) is used to create an \ac{OFDM}-modulated data signal to evaluate key performance metrics such as \ac{SNR} and data rate. The \ac{AC} signal produced by the \ac{AWG} is a peak-to-peak voltage of 1 V to prevent distortion or nonlinearities in the transmitter response. The discrete-time domain \ac{OFDM} signal is initiated by generating a pseudorandom binary stream. All \ac{OFDM} frames with bit and power allocation are generated with parameters set similar to those in \cite{ref37}. The modulated signals are combined with the \ac{DC} using a bias-tee (Mini-circuits, ZFBT-4R2GW+) and the output signal from the bias-tee is applied to an 847~nm VCSEL diode (TT Electronics, OPV314). Note that the wavelength of the \ac{VCSEL} falls within the peak responsivity region of the \ac{PPC}. In addition, the adaptive bit and energy loading algorithm is applied to dynamically adjust the number of bits and power assigned to each symbol based on channel conditions. This ensures efficient use of the modulation bandwidth, resulting in a higher achievable data rate.

The \ac{VCSEL} emitted a divergent beam with an optical power of 2.3~mW. This divergent laser beam is collimated using an aspheric condenser lens (Thorlabs, ACL7560U-B). The collimated beam is transmitted over a 1.5~m link length and is subsequently collected and focused onto a \ac{PPC} \ac{Rx} cell using a similar lens model. It is noted that the \ac{PPC} \ac{Rx} is enclosed in a metal cage to prevent \ac{RF} interference. The analogue \ac{OFDM} signal is captured with a simple \ac{Rx} circuit, which is the \ac{PPC} \ac{Rx} in parallel with a variable load resistor $R_L$ to define the operating point and maximize the available gain-bandwidth product. A variable resistor was tested over 0 to 1\,k$\Omega$ to identify the optimal impedance match near the short‑circuit condition. A load resistance ($R_L$) of 950\,$\Omega$ was selected to align with the circuit’s characteristic impedance, suppressing signal reflections. This configuration, operating near to short‑circuit, accelerates charge collection and supports the highest achievable data rate.

The electrical \ac{AC} signal at the receiver is processed using a spectrum analyser (Keysight, E4440A PSA) and an oscilloscope (Keysight, UXR0104B, 10~GHz) for the data communication measurements. This signal is then fed into an \ac{RF} amplifier (Mini-Circuits, ZHL-42W+) in the case of data communication. Hence, it enhances the signal-to-quantization-noise ratio to combat the inherent noise introduced by the oscilloscope's analog-to-digital conversion process. The amplified signal is then captured and converted to a digital signal using an oscilloscope and then sent to the \ac{PC} for decoding and further processing, see Fig.~\ref{fig:experimental_setup_photo}. The digital signal processing for \ac{OFDM} communication takes place in Matlab\textsuperscript{\textregistered} and is indicated in the \textit{OFDM communication} section below \cite{ref37}.

\subsection{OFDM communication}
\begin{figure*}[t!]
    \centering
    \begin{tikzpicture}
        \node[draw, line width=1pt, rectangle, inner sep=2pt, rounded corners=5pt] (figbox) {
            \begin{minipage}{\textwidth}
                \centering
                 \begin{subfigure}[b]{0.27\textwidth}
                    \centering
                    \includegraphics[width=\textwidth, trim=0cm 9cm 10cm 9cm, clip]{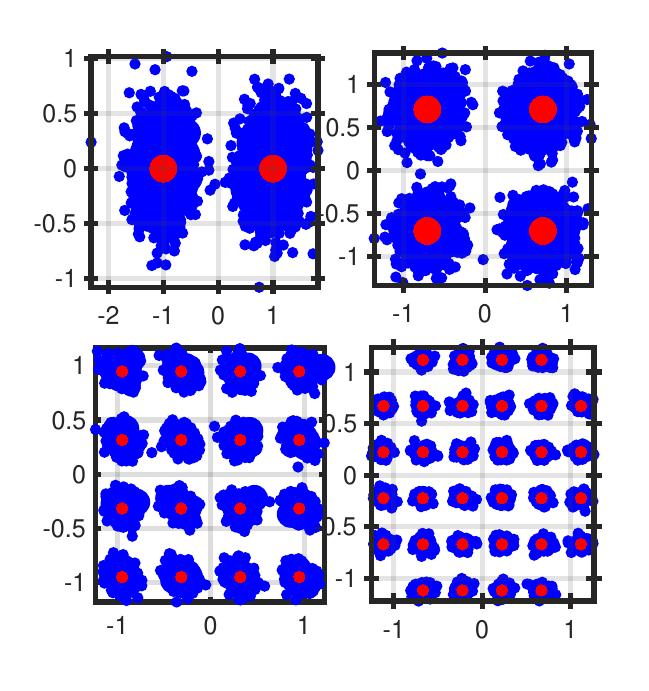}
                    \caption{}
                    \label{fig:subfig_c}
                \end{subfigure}
                \hfill
                 \begin{subfigure}[b]{0.24\textwidth}
                    \centering
                    \includegraphics[width=\textwidth, trim=0cm 0cm 4.5cm 0cm, clip]{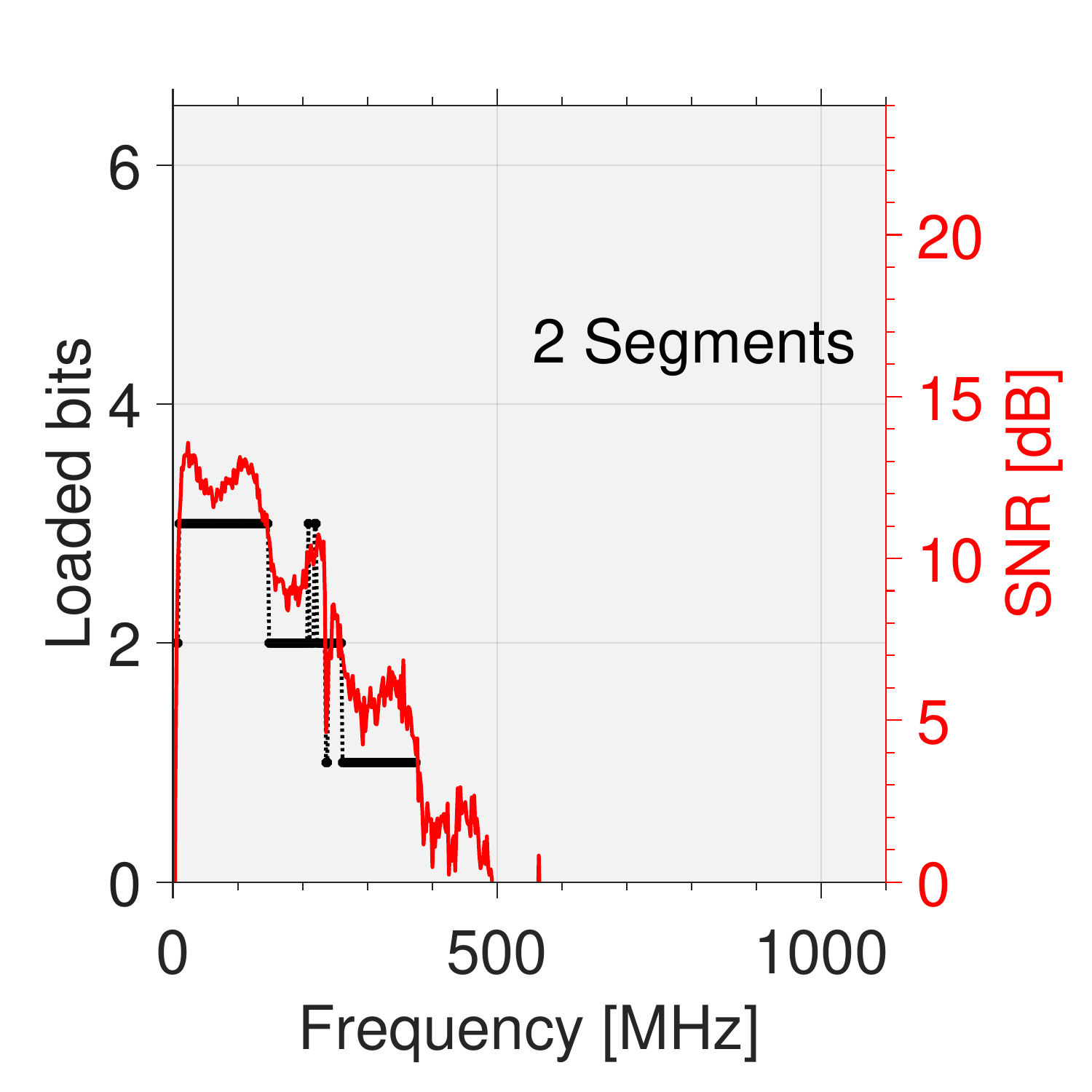}
                    \caption{}
                    \label{fig:2L_P}
                \end{subfigure}
                \hfill                
                \begin{subfigure}[b]{0.2\textwidth}
                    \centering
                    \includegraphics[width=\textwidth, trim=3.5cm 0cm 4.5cm 0cm, clip]{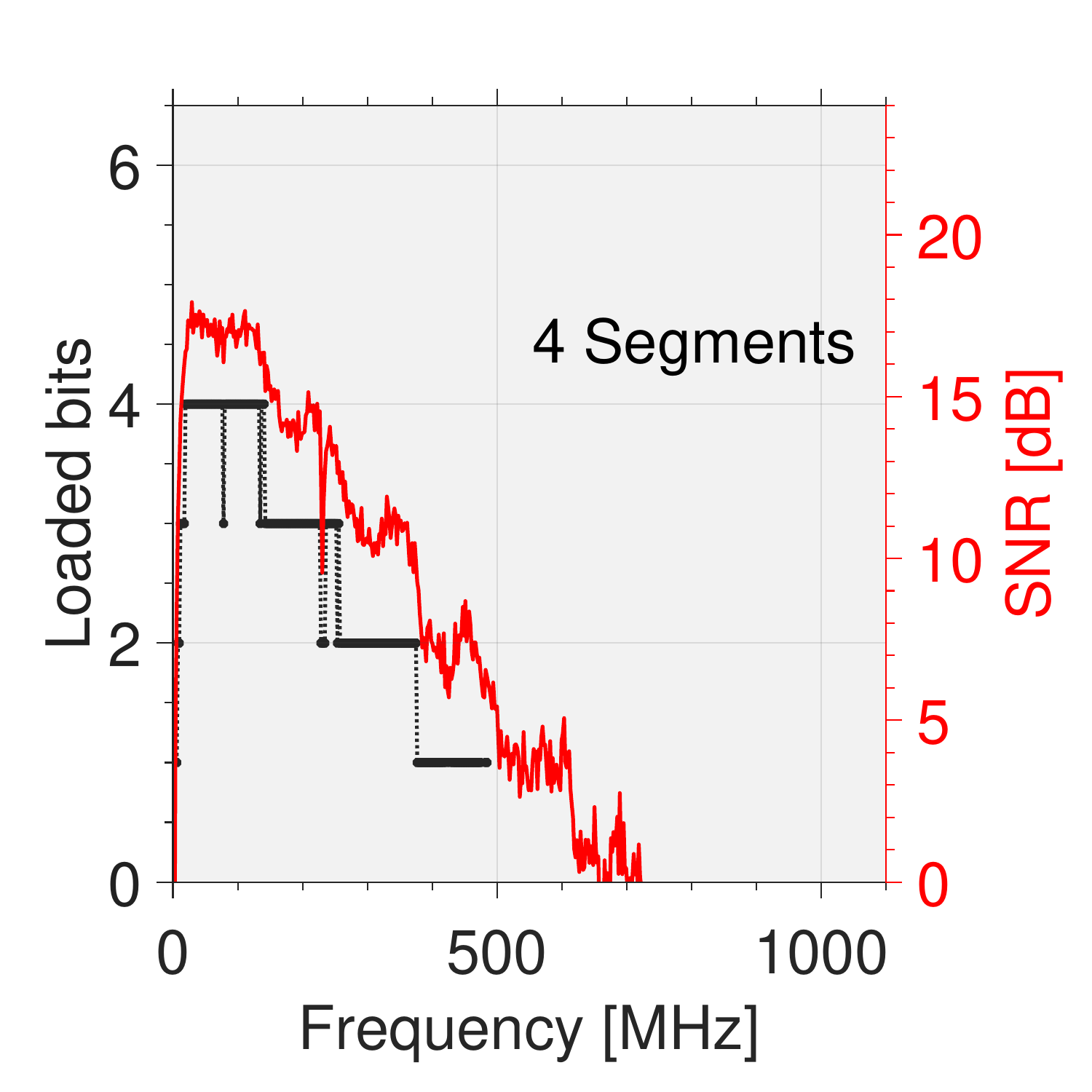}
                    \caption{}
                    \label{fig:4L_P}
                \end{subfigure}
                \hfill
                \begin{subfigure}[b]{0.25\textwidth}
                    \centering
                    \includegraphics[width=\textwidth, trim=3.5cm 0cm 0cm 0cm, clip]{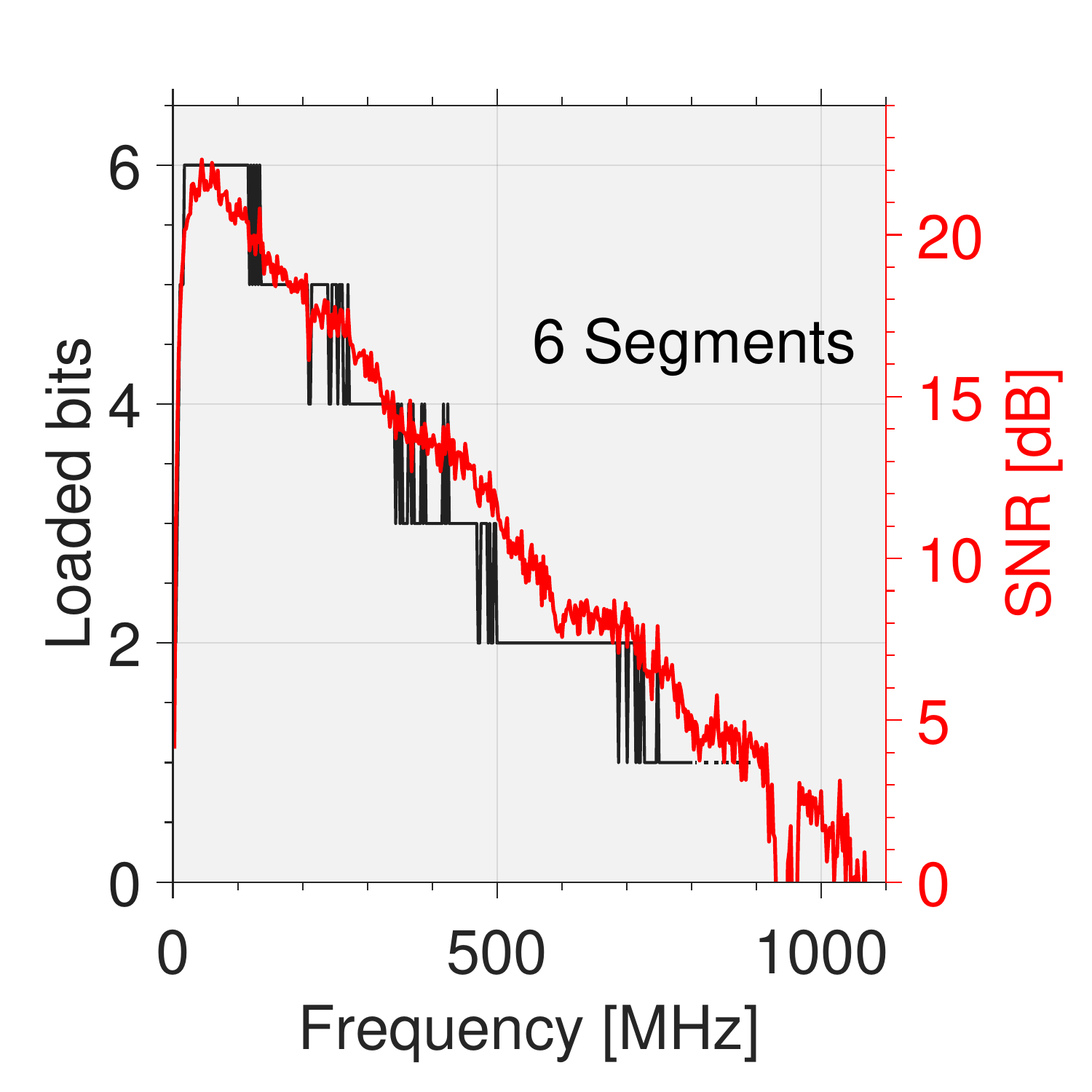}
                    \caption{}
                    \label{fig:6L_P}
                \end{subfigure}

            \end{minipage}
        };
    \end{tikzpicture}

    \caption{QAM symbol constellations and measured SNR performance of the PPC-based receiver: \textbf{(a)} Ideal transmitted symbols (red dots) and received symbols (blue dots) for QAM modulation orders $M = 2, 4, 16,$ and $32$; \textbf{(b--d)} measured \ac{SNR} of the \ac{PPC}-based receiver with adaptive bit allocation, implemented on a \ac{GaAs} device with a circular active area of diameter $d = 2.08~\mathrm{mm}$, for configurations with \textbf{(b)} 2 segments, \textbf{(c)} 4 segments, and \textbf{(d)} 6 segments.}
    \label{fig:estimated_snr_power}
\end{figure*}

The proposed communication link incorporates channel estimation and \ac{DCO}-\ac{OFDM} with \ac{QAM} schemes, which are implemented in software using a \ac{PC} running Matlab\textsuperscript{\textregistered} software. The creation of a discrete time domain \ac{OFDM} signal requires the following sequence \cite{ref6}: Pseudorandom bit generation, adaptive bit and power loading, M-\ac{QAM} modulation, \ac{IFFT}, oversampling and pulse shaping. The digital \ac{OFDM} frame is supplied to the \ac{AWG} that converts it to an analogue waveform. The analog \ac{OFDM} signal is captured in the receiver using the oscilloscope and converted to a digital signal. The separate stages of synchronization, matched filtering, down-sampling, \ac{FFT}, channel estimation, equalization and $M$-ary \ac{QAM} demodulation are used to process the discrete signal in Matlab\textsuperscript{\textregistered} \cite{ref6}. The application of Hermitian symmetry before the \ac{IFFT} operation is needed for the generation of a real-valued \ac{OFDM} signal \cite{ref5}. Thus, while a \ac{FFT} size of 1024 is used, only 511 subcarriers transfer unique information. The rest of the OFDM parameters of the system are as follows: a cyclic prefix length of $5$ samples, optimal clipping thresholds of $-3.2\sigma_x$ and $3.2\sigma_x$, respectively, where $\sigma_x$ denotes the standard deviation of the time-domain OFDM signal $x$ (see \cite{ref6}), and a maximum QAM constellation size of $1024$ used for adaptive bit and power loading; see Figure \ref{fig:subfig_c}.

Figure \ref{fig:estimated_snr_power}~(b, c, and d) illustrates the estimated \ac{SNR} (red curve, right y-axis) and adaptive bit loading (black steps, left y-axis) at \ac{BER} thresholds of $6.6\times10^{-4}$  for \ac{GaAs} device with a circular active area of diameter \( d = 2.08~\mathrm{mm} \) and configurations with segmentations of $2$, $4$, and $6$ . 
Increased segmentation enhances \ac{SNR} across a broader frequency range, allowing more bits to be allocated to each subcarrier and thereby enhancing the overall data rate. The 2-segment configuration exhibits limited performance, with \ac{SNR} peaks at $12$~dB and supports a maximum of $3$~bits per subcarrier (see Fig.~\ref{fig:2L_P}). The 4-segment configuration depicted in Fig.~\ref{fig:4L_P} shows a moderate improvement, reaching \ac{SNR} levels up to $17$~dB and enabling up to 4~bits. The 6-segment configuration achieves the highest efficiency, with \ac{SNR} exceeding 20~dB in low-frequency regions and supporting more than 6~bits per subcarrier, see Fig.~\ref{fig:6L_P}. These results demonstrate the critical role of segmentation in improving signal integrity and optimizing spectral utilization.

\section{Results and Discussion}

\subsection{Experimental Results}
\begin{figure*}[t!]
    \centering
    \begin{tikzpicture}
    \node[draw, line width=1pt, rectangle, inner sep=4pt, rounded corners=5pt] (figbox) {
        \begin{minipage}{\textwidth}
        \centering
        \begin{tabular}{@{}p{0.55\textwidth}@{\hspace{0.02\textwidth}}p{0.43\textwidth}@{}}
            \begin{subfigure}{\linewidth}
                \centering
                \includegraphics[width=\linewidth, height=8cm, keepaspectratio, trim=0cm 0cm 0cm 0cm, clip]{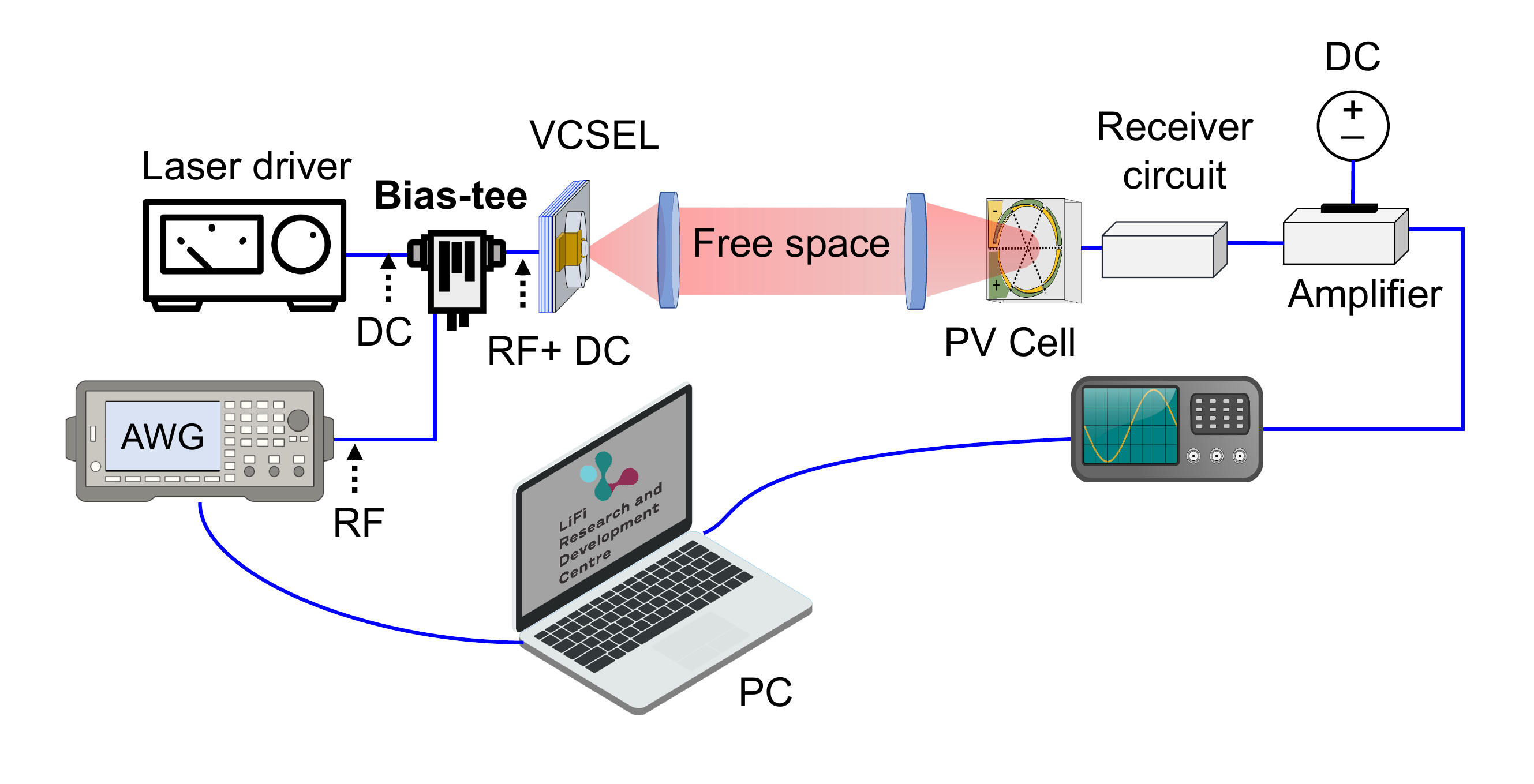}
                \caption{}
                \label{fig:BlockDiagram}
            \end{subfigure}
            &
            \begin{subfigure}{\linewidth}
                \centering
                \includegraphics[width=\linewidth, height=8cm, keepaspectratio, trim=2.5cm 0cm -2cm 0cm, clip]{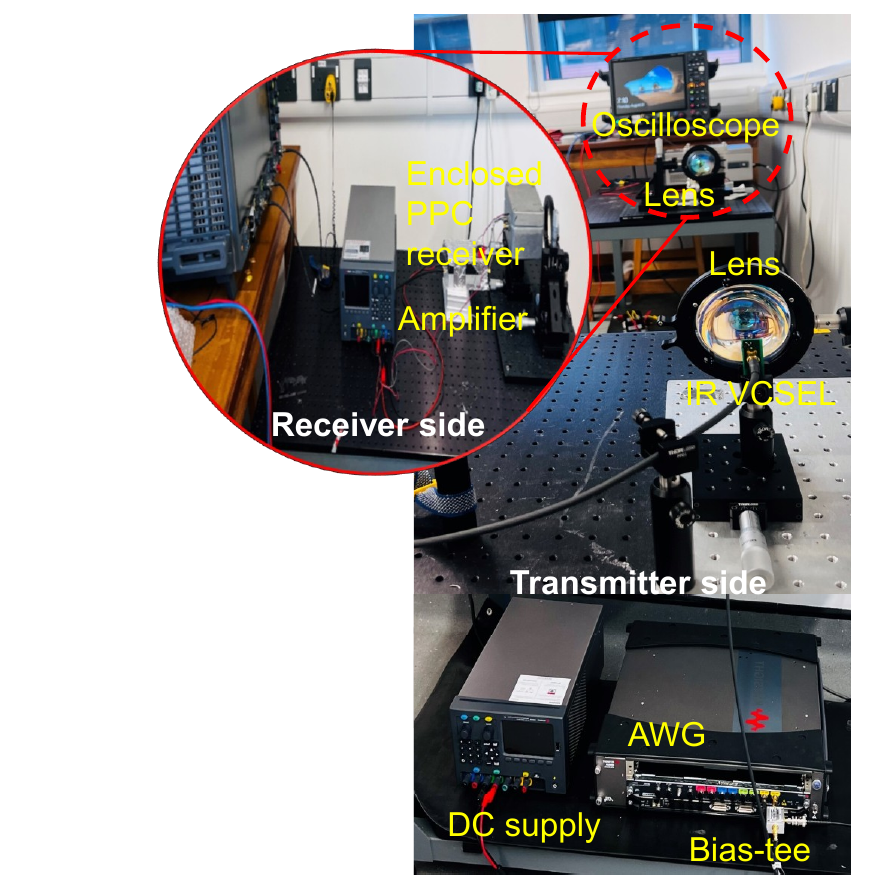}
                \caption{}
                \label{fig:experimental_setup_photo}
            \end{subfigure}
            \\
            \begin{subfigure}{\linewidth}
                \centering
                \includegraphics[width=\linewidth, keepaspectratio, trim=0cm 7cm 0cm 6cm, clip]{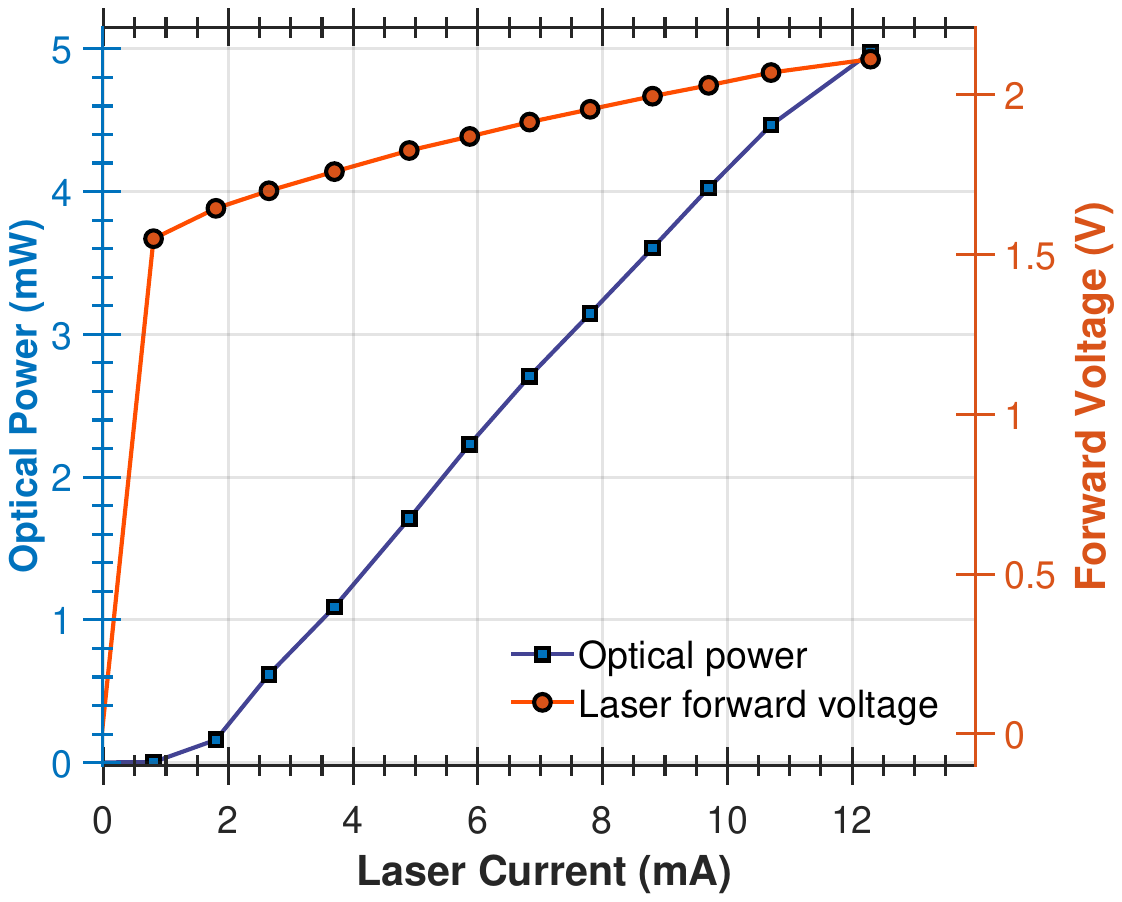}
                \caption{}
                \label{fig:Tx_Charactristics}
            \end{subfigure}
            &
            \begin{subfigure}{\linewidth}
                \centering
                \includegraphics[width=0.4\linewidth, trim=3.5cm 0cm 0cm 0cm, clip]{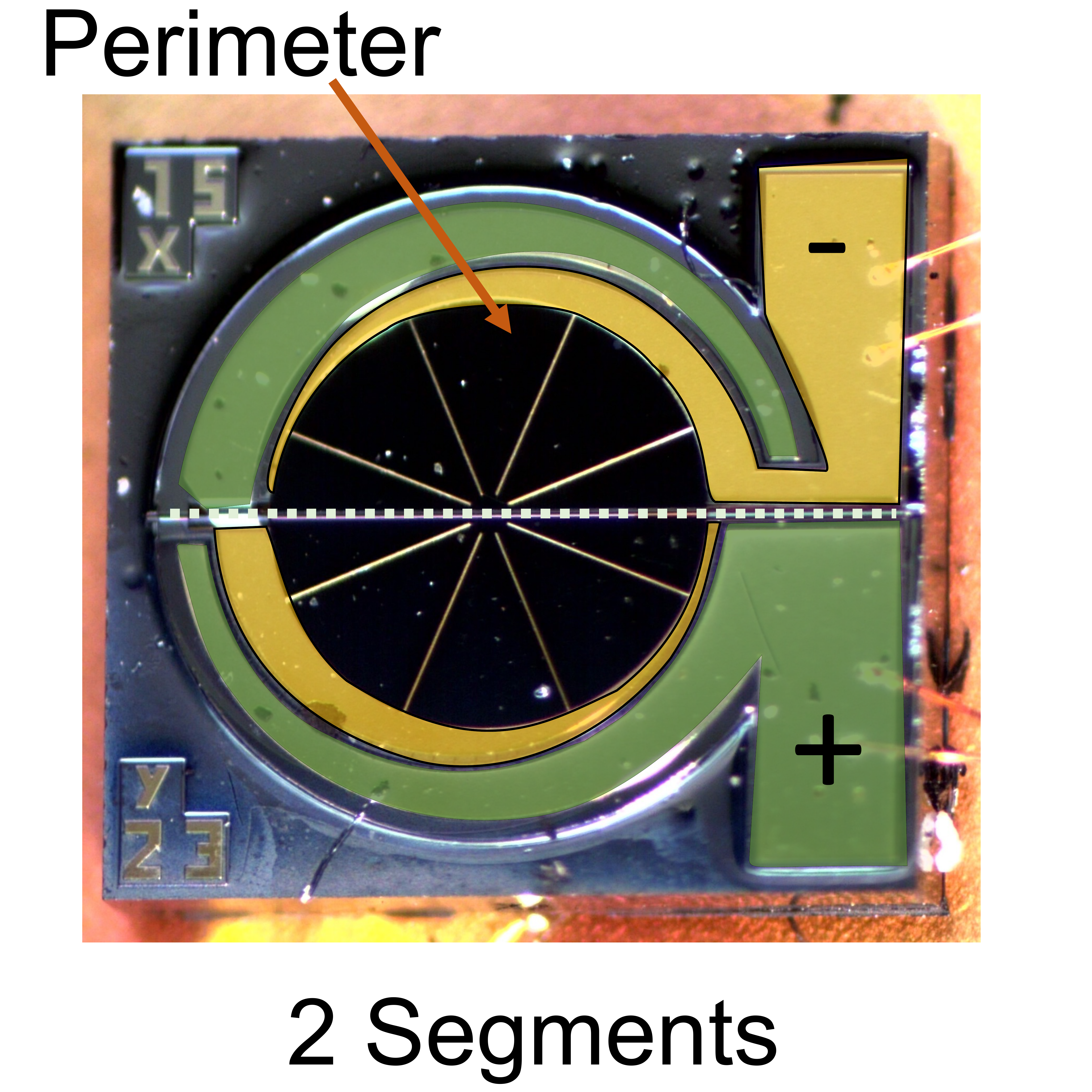}\hfill
                \includegraphics[width=0.4\linewidth, trim=4cm 0cm 0cm 0cm, clip]{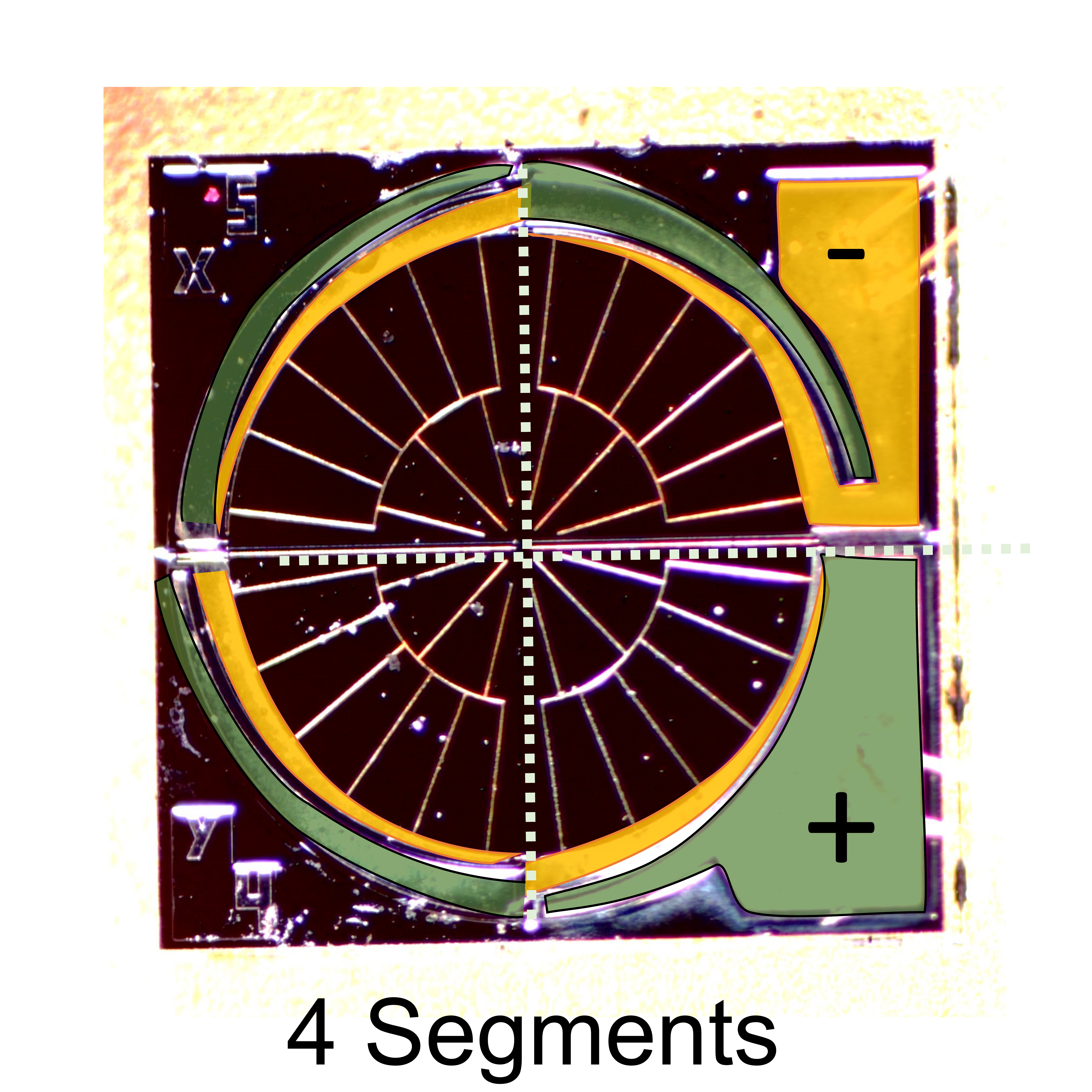}\hfill
                \includegraphics[width=0.4\linewidth, trim=4cm 0cm 0cm 0cm, clip]{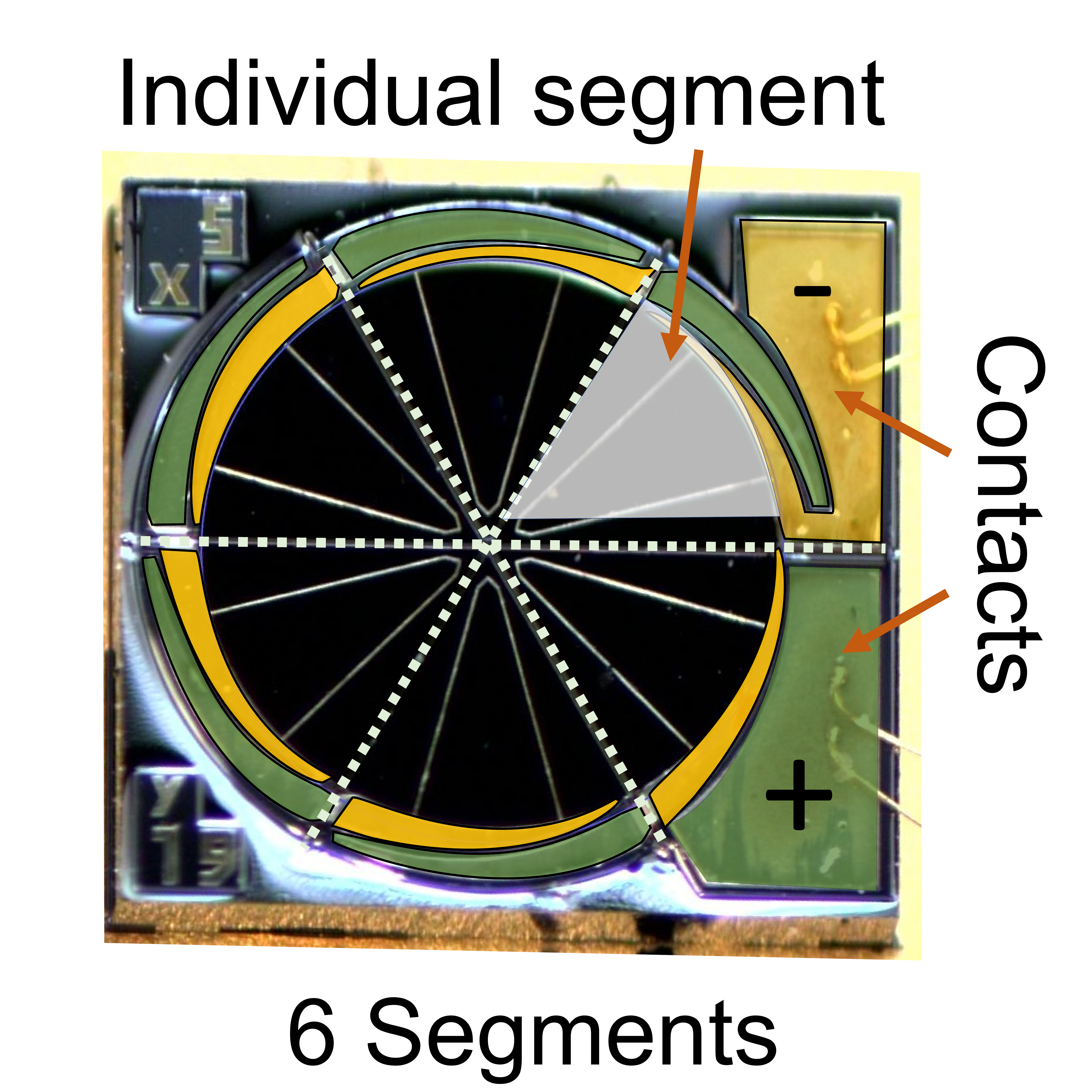}
                \caption{}
                \label{fig:Samplec}
            \end{subfigure}
        \end{tabular}
        \end{minipage}
    };
    \end{tikzpicture}
    \caption{Experimental setup and characteristics: (a) Block diagram for the proposed \ac{GaAs}-based \ac{PPC} receiver system; (b) Photograph of the experimental setup of the eye-safe infrared wireless communication system with a link length of 1.5 meters; (c) Measured characteristics of the \ac{VCSEL} laser source; (d) Colored photographs of example devices. Circular sectors are interconnected outside the circular active area across isolation trenches (white dashed lines) from terminals underneath the active junction (green) to front-side contacts (yellow). Larger pads at one side (“+” and “–“) enable electrical contacting of the entire series-connected string via thin wire bonding.}
    \label{fig:experimental_setup}
\end{figure*}

We have developed an eye‑safe infrared wireless communication system with a link length of 1.5 meters, based on a multi‑segment \ac{GaAs}-based \ac{PPC} receiver that harvests energy and communicates data. Figures \ref{fig:BlockDiagram} and \ref{fig:experimental_setup_photo} illustrate the system architecture and a photograph of the experimental setup, respectively. A detailed description of the optical and electronic components is provided in the \textit{Experimental setup} section. On the transmitter side, the peak-to-peak voltage of the \ac{AWG} and the \ac{DC} bias current are carefully tuned to ensure operation within the linear region of the \ac{VCSEL}, thereby maximizing the average \ac{SNR} across the modulation bandwidth. The transmitted optical power is measured at 2.3~mW, see Fig.~\ref{fig:Tx_Charactristics}. Details on compliance with eye safety standards are provided in the \textit{Appendix} section.

On the receiver side, the \ac{PPC} is investigated in terms of size and number of segments, we optimize the trade‑off between harvested power and high‑speed data detection. These design choices are based on our previous work \cite{ref32} and allow to study the influence of junction capacitance of the subcells, which scales with area, and the series connection. A sample of the \ac{GaAs}-based \ac{PPC} is indicated in Fig. \ref{fig:Samplec} and the design methodology is detailed in \textit{Sample Description} section. The communication channel is assessed using \ac{SNR}, data rate, and \ac{BER} across various operating points of the evaluated sample. The operating points are derived from the \ac{I-V} curve measurements for each sample and obtained using a parallel load resistance $R_L$. Setting the \ac{PPC} resistive load close to the short‑circuit condition, with $R_L$ of $950~\Omega$ chosen for optimal impedance matching, minimises parasitic capacitance and accelerates charge collection, thereby allowing increased bandwidth for high‑speed operation \cite{ref22}. Given that the primary objective of these experiments was data transmission, no active homogenization of the incoming optical beam was performed, despite the well-established sensitivity of \acp{MIM} devices to nonuniform illumination \cite{WAGNER2017287}.

\begin{figure*}
    \centering
    \begin{tikzpicture}
        \node[draw, line width=1pt, rectangle, inner sep=2pt, rounded corners=5pt] (figbox) {    
            \begin{minipage}{0.98\textwidth}
                \centering
            \begin{tikzpicture}
            \vspace {1em}
                \matrix [matrix of nodes, column sep=1em, nodes={anchor=west}] {
                    \node[draw, circle, fill=blue, minimum size=6pt, inner sep=0pt] {}; & {Two segments} &
                    \node[draw, regular polygon, regular polygon sides=3, fill=red!70!black, rotate=0, minimum size=10pt, inner sep=0pt, rounded corners=0pt] {}; &{Four segments} &
                    \node[draw, rectangle, fill=yellow, minimum size=8pt, inner sep=0pt, rounded corners=0pt] {}; & {Six segments} \\
                };
            \end{tikzpicture}
                \begin{subfigure}[b]{\textwidth}
                    \centering
                    \includegraphics[width=0.35\textwidth]{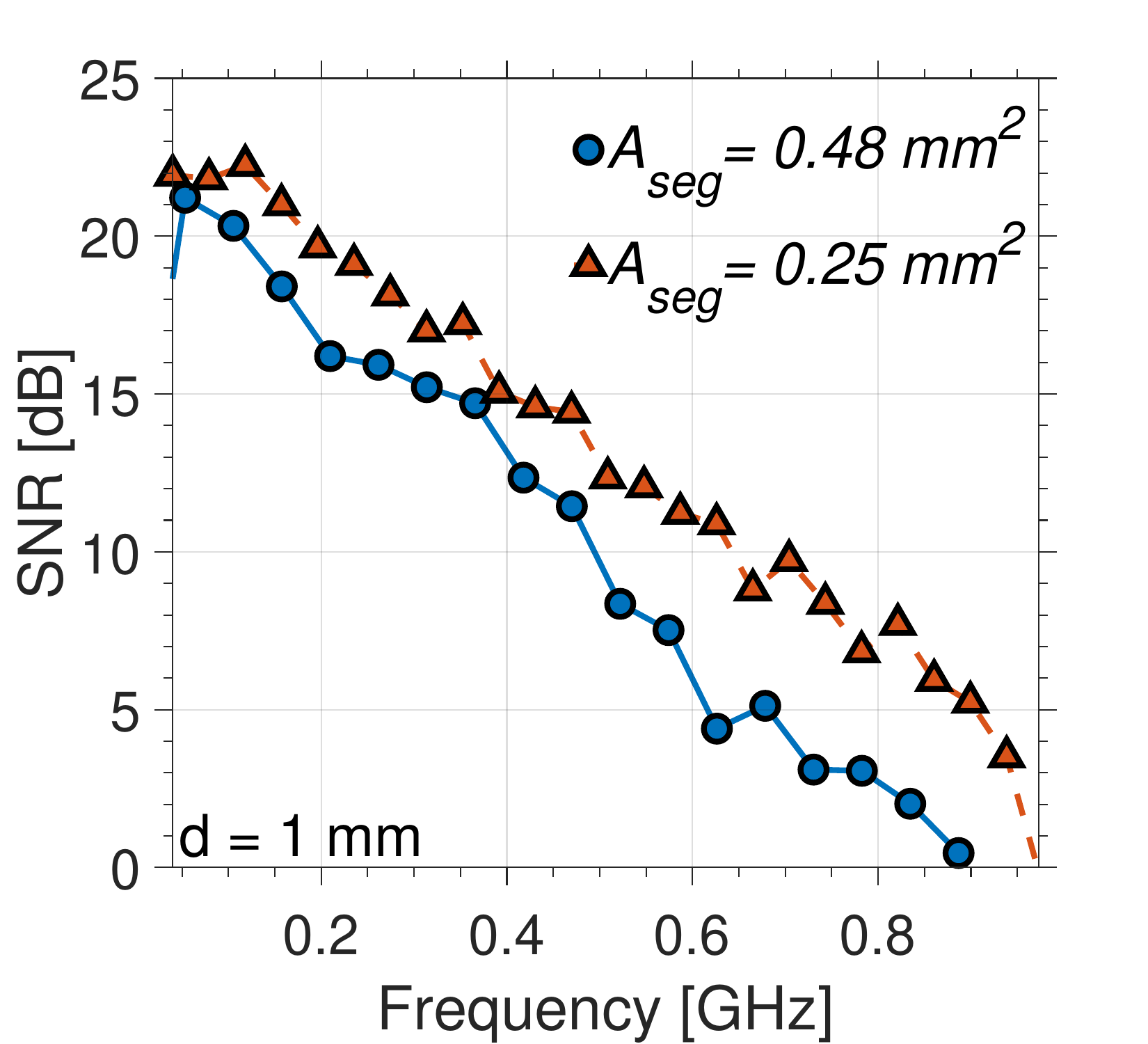}
                    \includegraphics[width=0.3\textwidth, trim=4cm 0cm 0cm 0cm, clip]{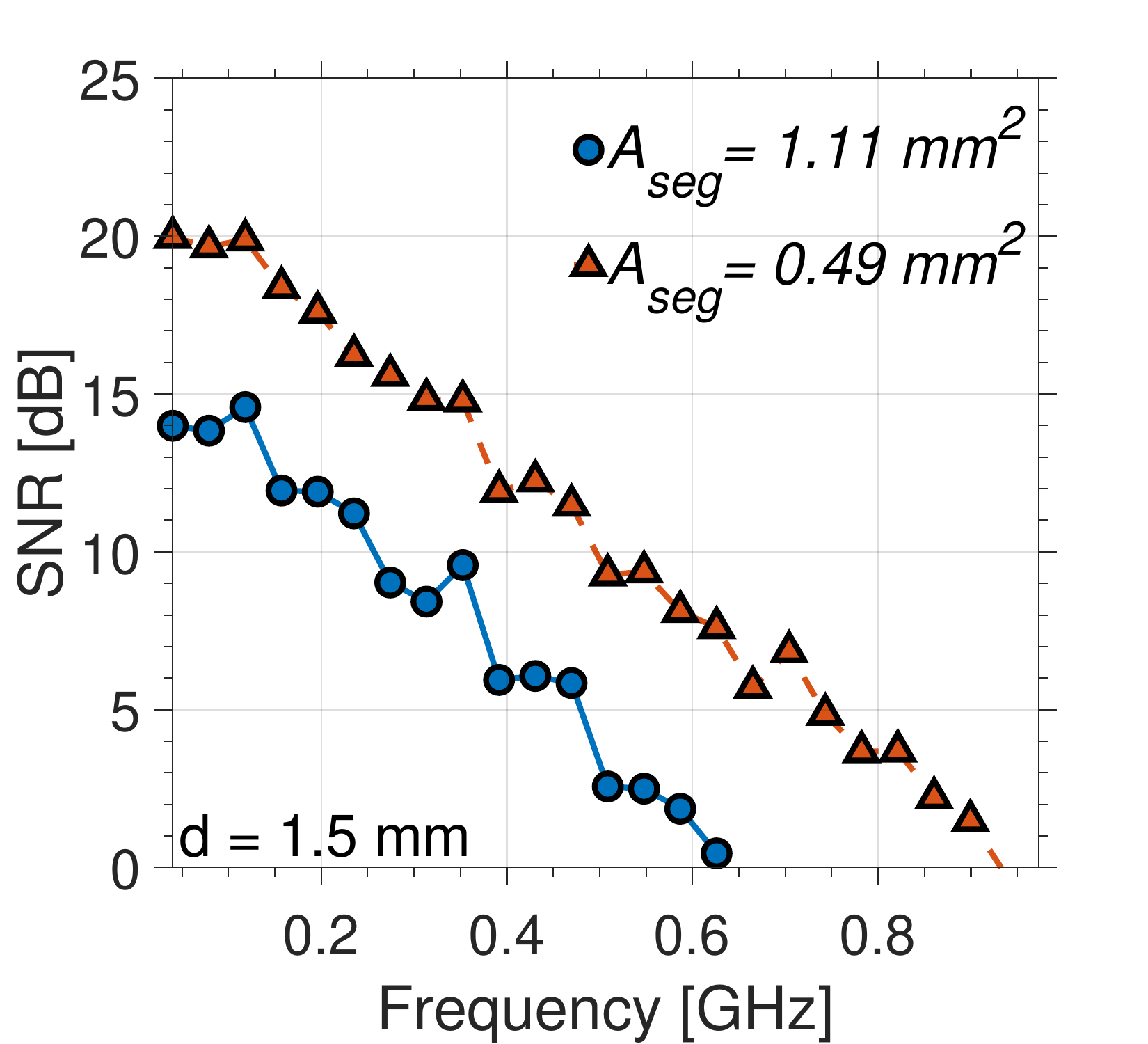}
                    \includegraphics[width=0.3\textwidth, trim=4cm 0cm 0cm 0cm, clip]{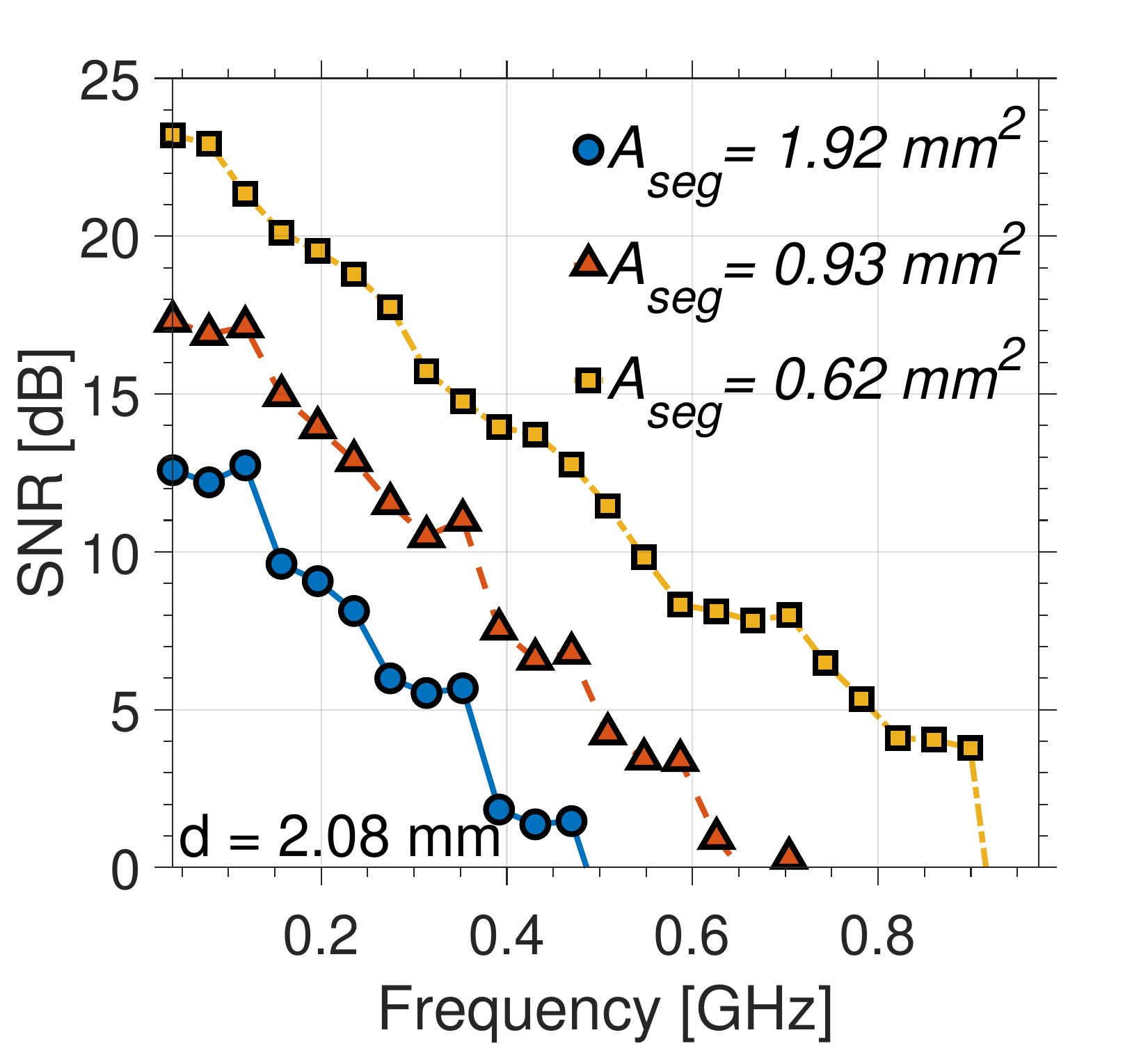}
                    \caption{}
                    \label{fig:SNR_nature_quality}
                \end{subfigure}
                \begin{subfigure}[b]{\textwidth}
                    \centering
                    \includegraphics[width=0.365\textwidth]{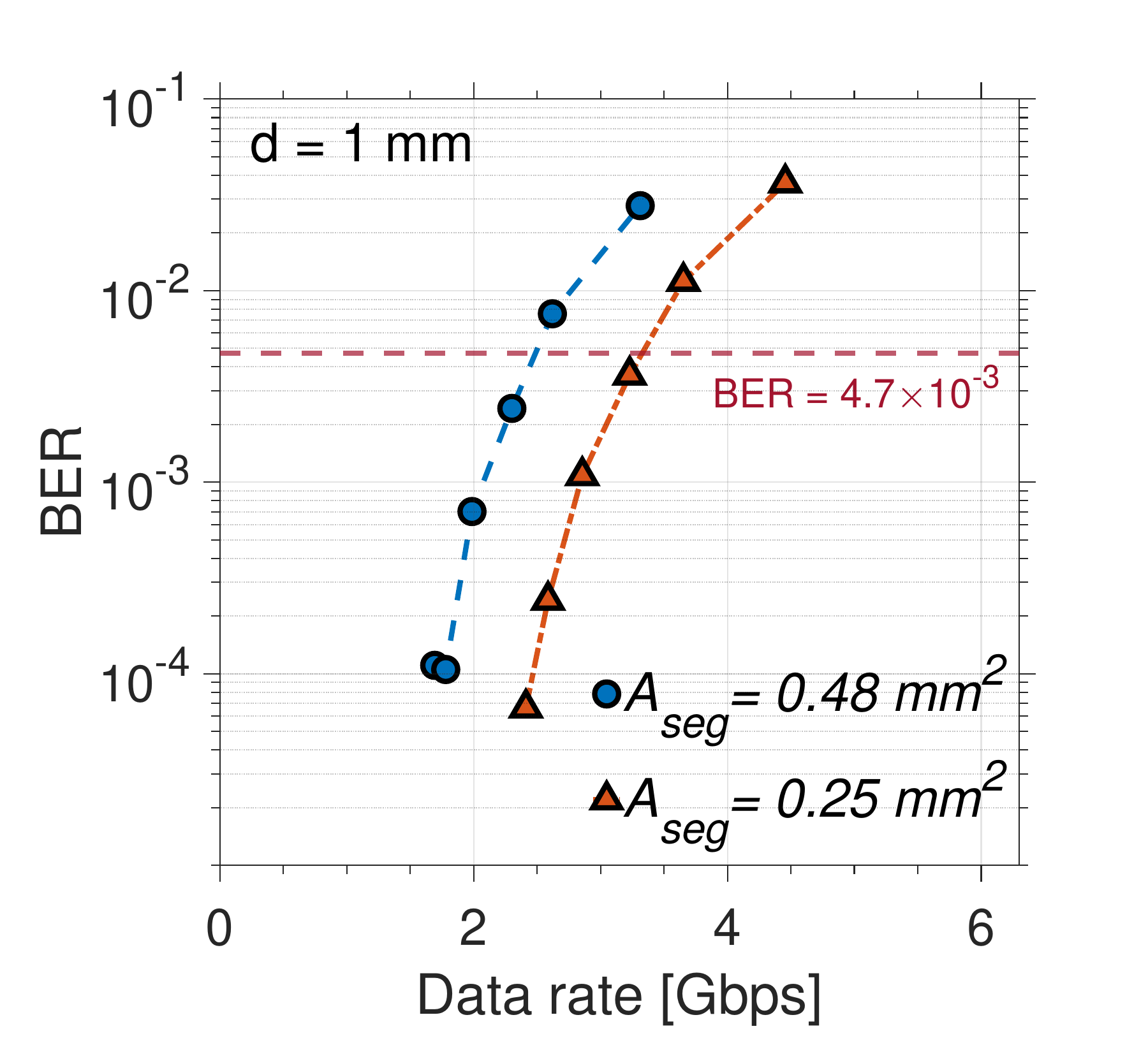}
                    \includegraphics[width=0.3\textwidth, trim=5cm 0cm 0cm 0cm, clip]{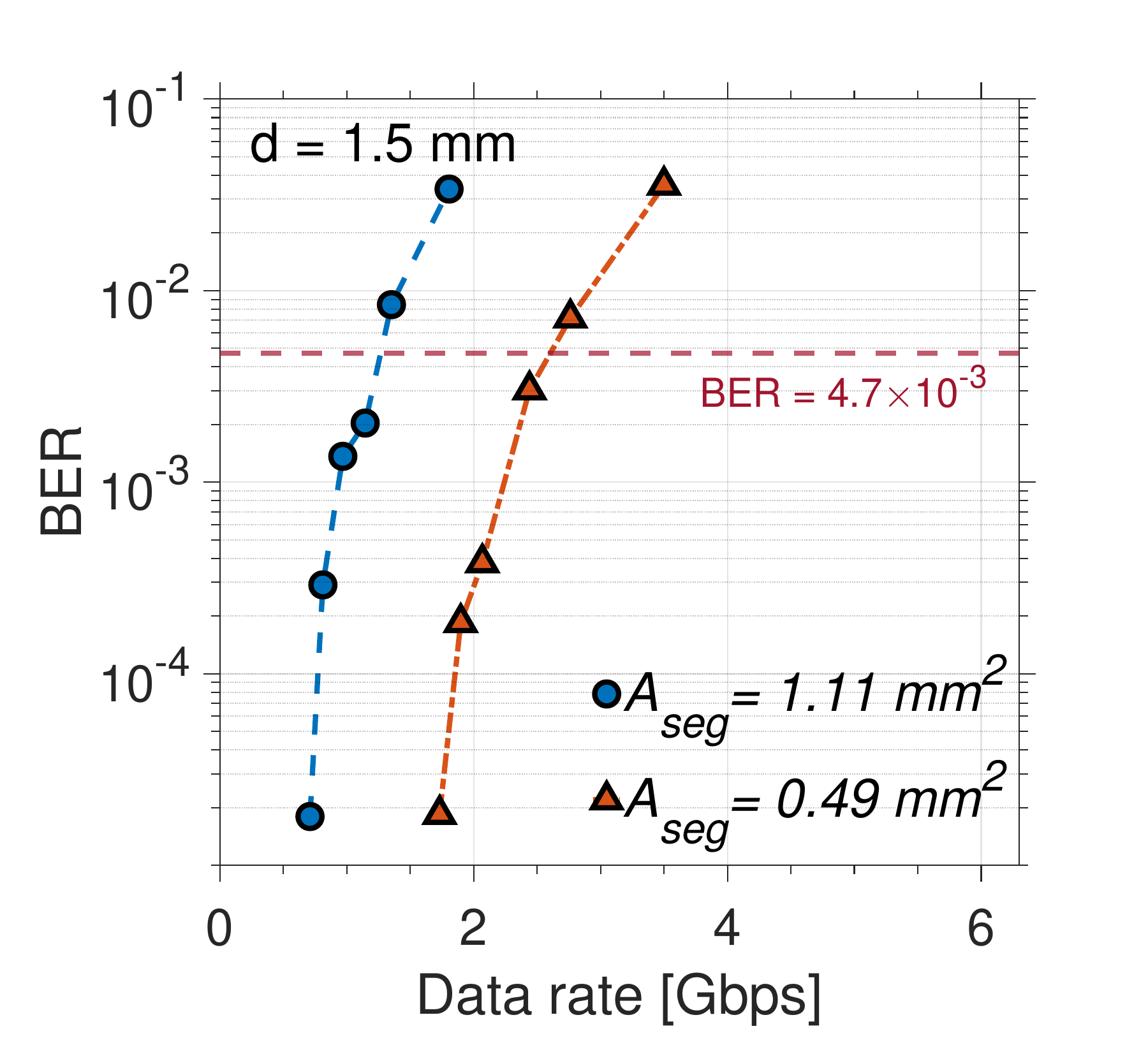}
                    \includegraphics[width=0.3\textwidth, trim=5cm 0cm 0cm 0cm, clip]{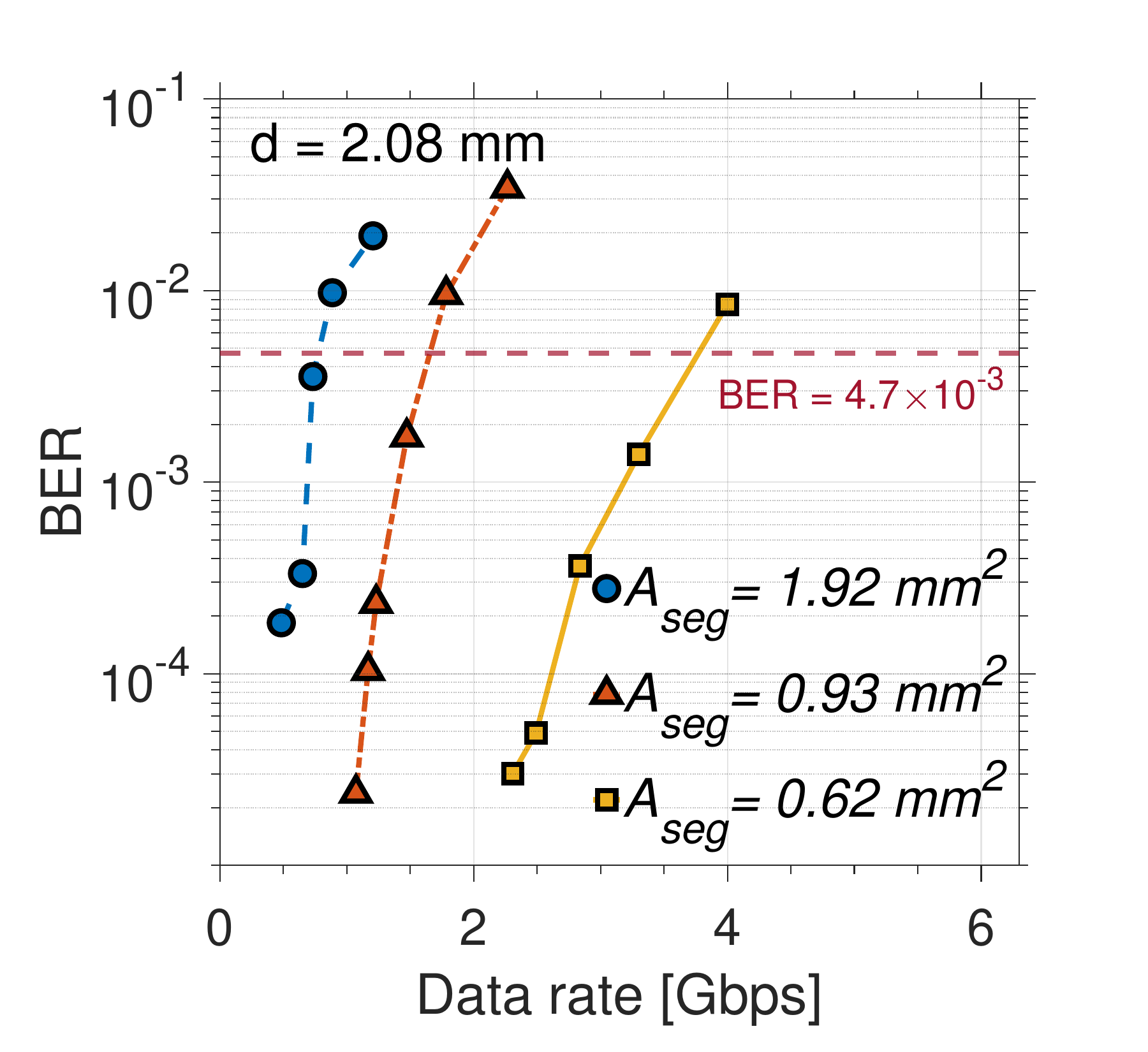}
                    \caption{}
                    \label{fig:BER_nature_quality}
                \end{subfigure}

                         \begin{subfigure}[b]{\textwidth}
                    \centering
                    \includegraphics[width=0.358\textwidth]{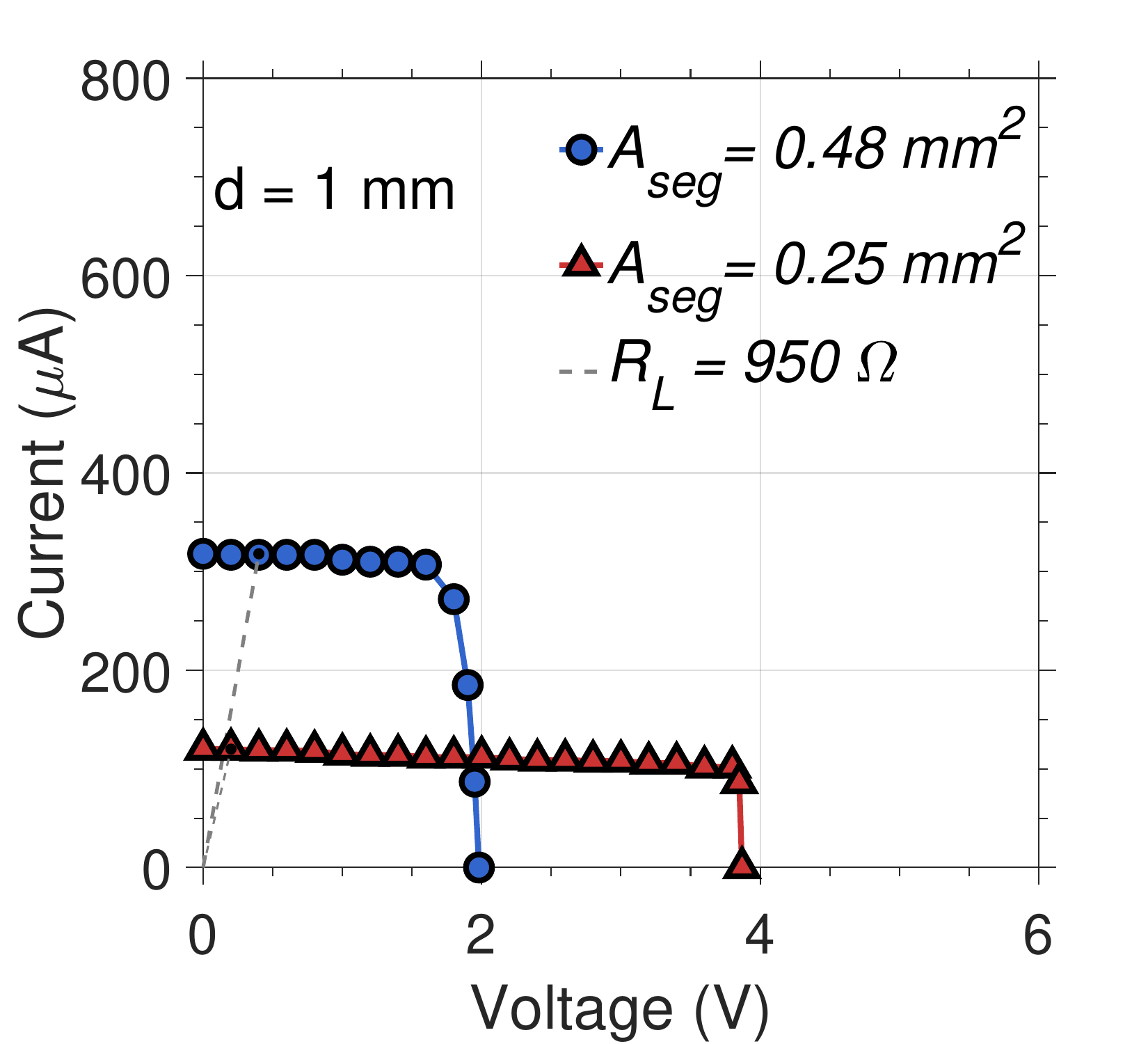}
                    \includegraphics[width=0.3\textwidth, trim=4.5cm 0cm 0cm 0cm, clip]{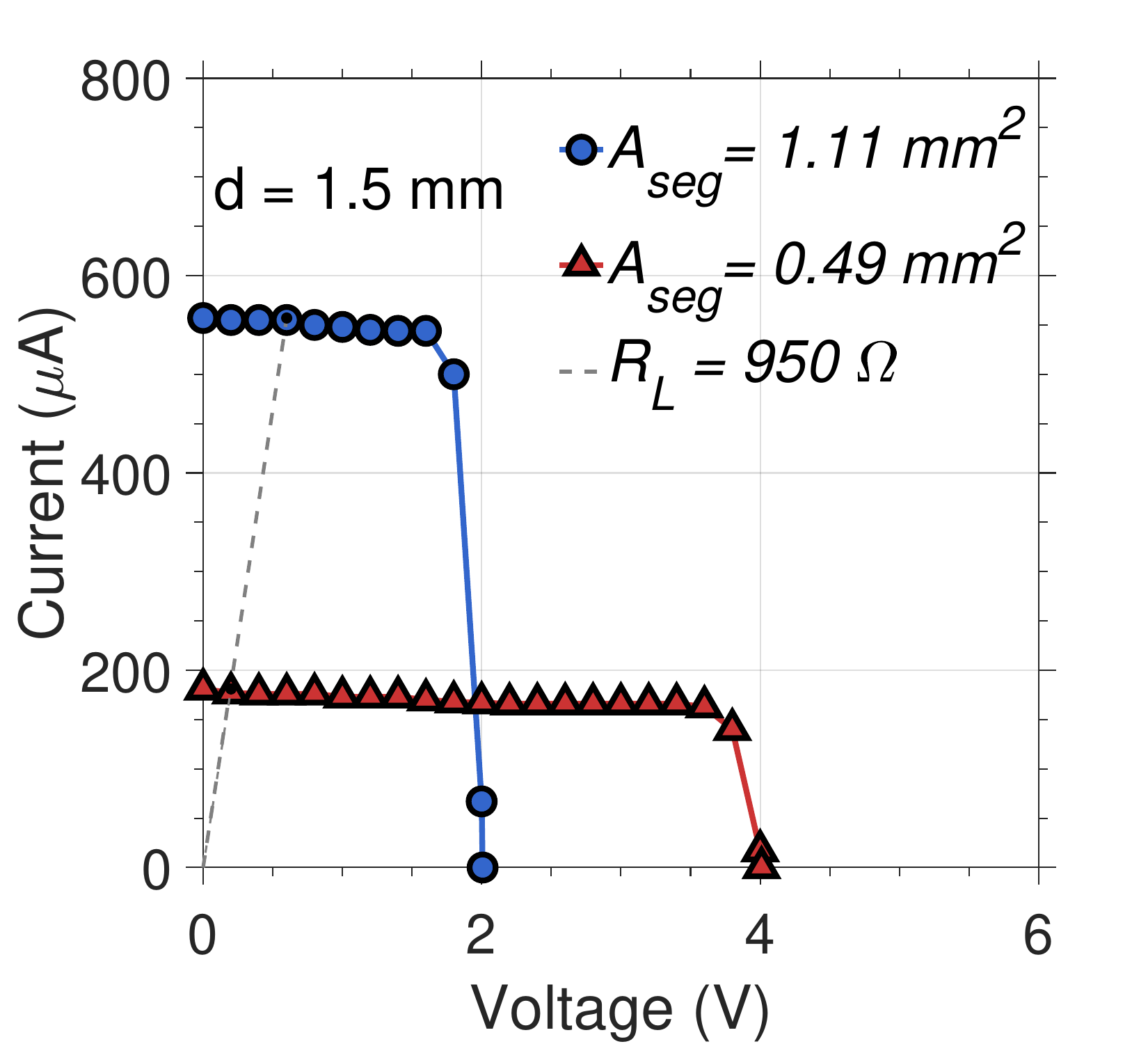}
                    \includegraphics[width=0.3\textwidth, trim=4.5cm 0cm 0cm 0cm, clip]{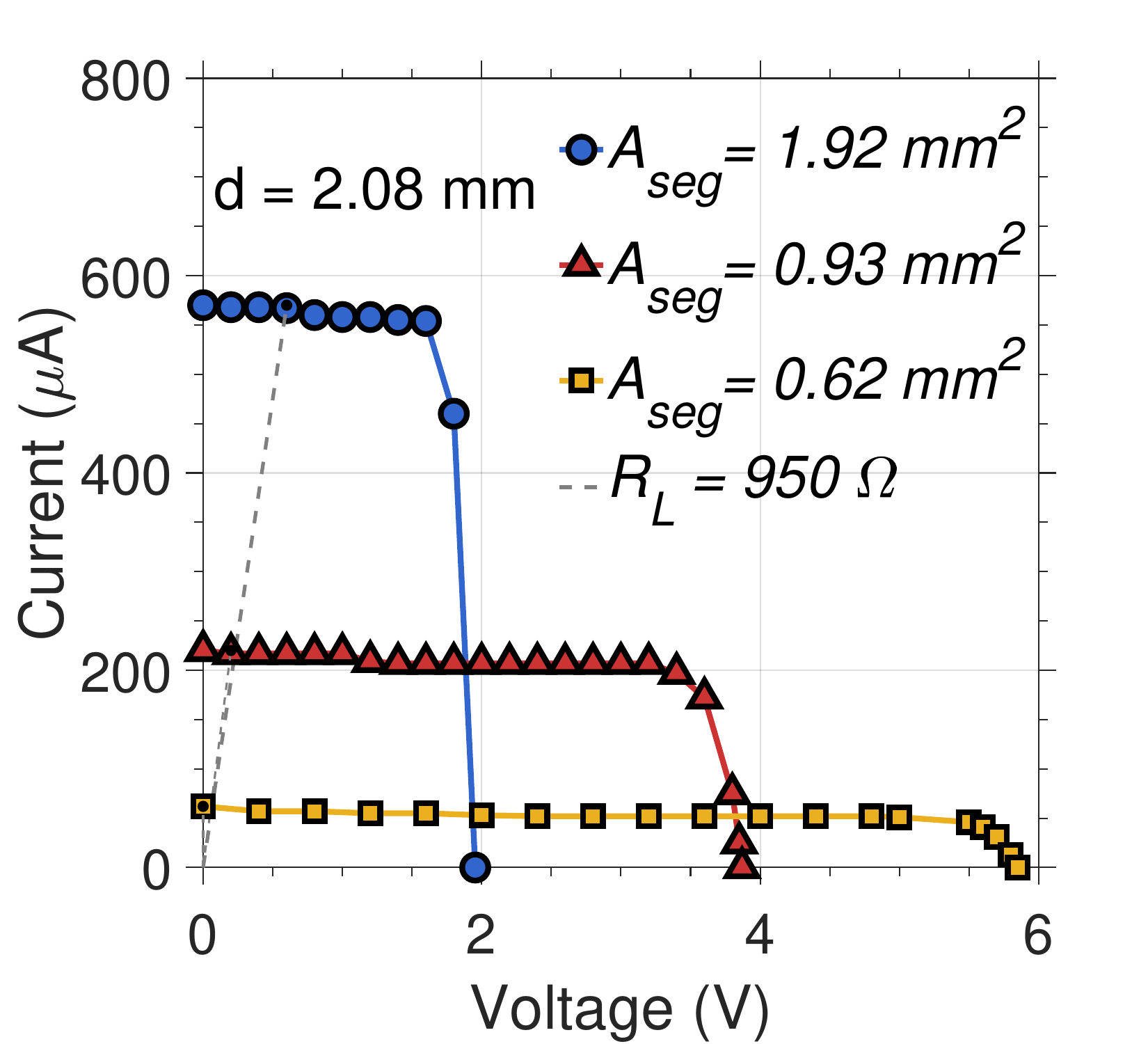}
                    \caption{}
                    \label{fig:Fig_IV_Nature}
                \end{subfigure}
            \end{minipage}
        };
    \end{tikzpicture}

    \caption{Recorded measurements for the \ac{PPC}-based \ac{GaAs} receiver. The receiver features circular active areas with diameters $d$ of 1~mm, 1.5~mm, and 2.08~mm. The figure presents: \textbf{a.} measured \ac{SNR} for various numbers of segments; \textbf{b.} \ac{BER} results versus data rate for different segment counts; and \textbf{c.} \ac{PPC} \ac{I-V} curve measured using a transmitter with an emitted optical power of $2.3$~mW, a \ac{DC} bias of 1.78~V and a current of 6~mA.}
   \label{fig:2} 
\end{figure*}

Figure \ref{fig:SNR_nature_quality} shows the \ac{SNR} and communication bandwidth of the proposed system, operating near the short-circuit condition, which are evaluated using segmented \ac{GaAs} \ac{PV} cells with diameters of 1\,mm (S), 1.5\,mm (M), and 2.08\,mm (L), corresponding to cell areas of $0.79~\text{mm}^2$, $1.77~\text{mm}^2$, and $3.40~\text{mm}^2$, respectively. The cells are segmented into 2, 4, or 6 subcells, with segment areas ranging from $0.25~\text{mm}^2$ to $1.92~\text{mm}^2$. The communication bandwidth, defined as the frequency range where \ac{SNR} remains above 0~dB, increased with the number of segments.

For the 1~mm cell, bandwidth increased from 0.88~GHz with 2 segments (shown in the blue line) to 0.94~GHz with 4 segments of subcells  (shown in the red line). The 1.5~mm cell shows an increase from 0.62~GHz to 0.93~GHz for 2 and 4 segments, respectively. Note that  6 segment configurations were developed for the cell variants ($d$ = 1 mm and 1.5 mm), however, their comprehensive data are not reported here due to challenges encountered during measurement, inconsistent mounting stability and suboptimal sample uniformity, which affected result reliability. The 2.08~mm cell achieves bandwidths of 0.49~GHz, 0.66~GHz, and 0.96~GHz for 2, 4, and 6 segments, respectively, peaking at 0.96~GHz with 6 segments, demonstrating enhanced performance with finer segmentation; see Table~\ref{tab:performance_metrics}.

\begin{table*}[!]
\centering
\caption{Recorded performance metrics of the proposed system using the segmented GaAs photovoltaic cells across various configurations and a VCSEL transmitter with an optical power of 2.3~mW.}
\label{tab:performance_metrics}
\begin{tabular}{p{8cm}ccccccc} 
\toprule
\textbf{\ac{GaAs} Cell diameter [mm]} & \multicolumn{2}{c}{\textbf{1 (S)}} & \multicolumn{2}{c}{\textbf{1.5 (M)}} & \multicolumn{3}{c}{\textbf{2.08 (L)}} \\
\textbf{\ac{GaAs} Cell size [mm\textsuperscript{2}]} & \multicolumn{2}{c}{\textbf{0.785}} & \multicolumn{2}{c}{\textbf{1.767}} & \multicolumn{3}{c}{\textbf{3.397}} \\

\cmidrule(lr){2-3} \cmidrule(lr){4-5} \cmidrule(lr){6-8}
\textbf{Number of segments} & 2 & 4 & 2 & 4 & 2 & 4 & 6  \\
\midrule
\textbf{Segment (junction) area [mm$^2$]} & 0.48 & 0.25 & 1.11 & 0.49 & 1.92 & 0.93 & 0.62 \\
\textbf{Power harvested at MPP current [mW]} & 0.49 & 0.39 & 0.89 & 0.59 & 0.89 & 0.67 & 0.23\\

\bottomrule
\end{tabular}
\end{table*}
 
  A \ac{BER} threshold of $4.7 \times 10^{-3}$ according to a staircase \ac{FEC} code with 6.25\% overhead \cite{Zhang2014}. The recorded data rate at this level ranges from 0.761~Gbps to 3.8~Gbps across the same cell configurations. Specifically, the 1~mm cell's data rate increases from 2.44~Gbps (2 segments) to 3.29~Gbps (4 segments), while the 1.5~mm cell advances from 1.23~Gbps (2 segments) to 2.56~Gbps (4 segments). The 2.08~mm cell increases from 0.761~Gbps (2 segments) to 1.59~Gbps (4 segments) and reaches 3.8~Gbps with 6 segments, as depicted in \ref{fig:BER_nature_quality}. This maximum recorded data rate is four times higher than previously reported values at an operation point near to a short circuit. 
 
\begin{figure}[b!]
    \centering
    \includegraphics[width=1\columnwidth, trim=0cm 0cm 0cm 0cm, clip]{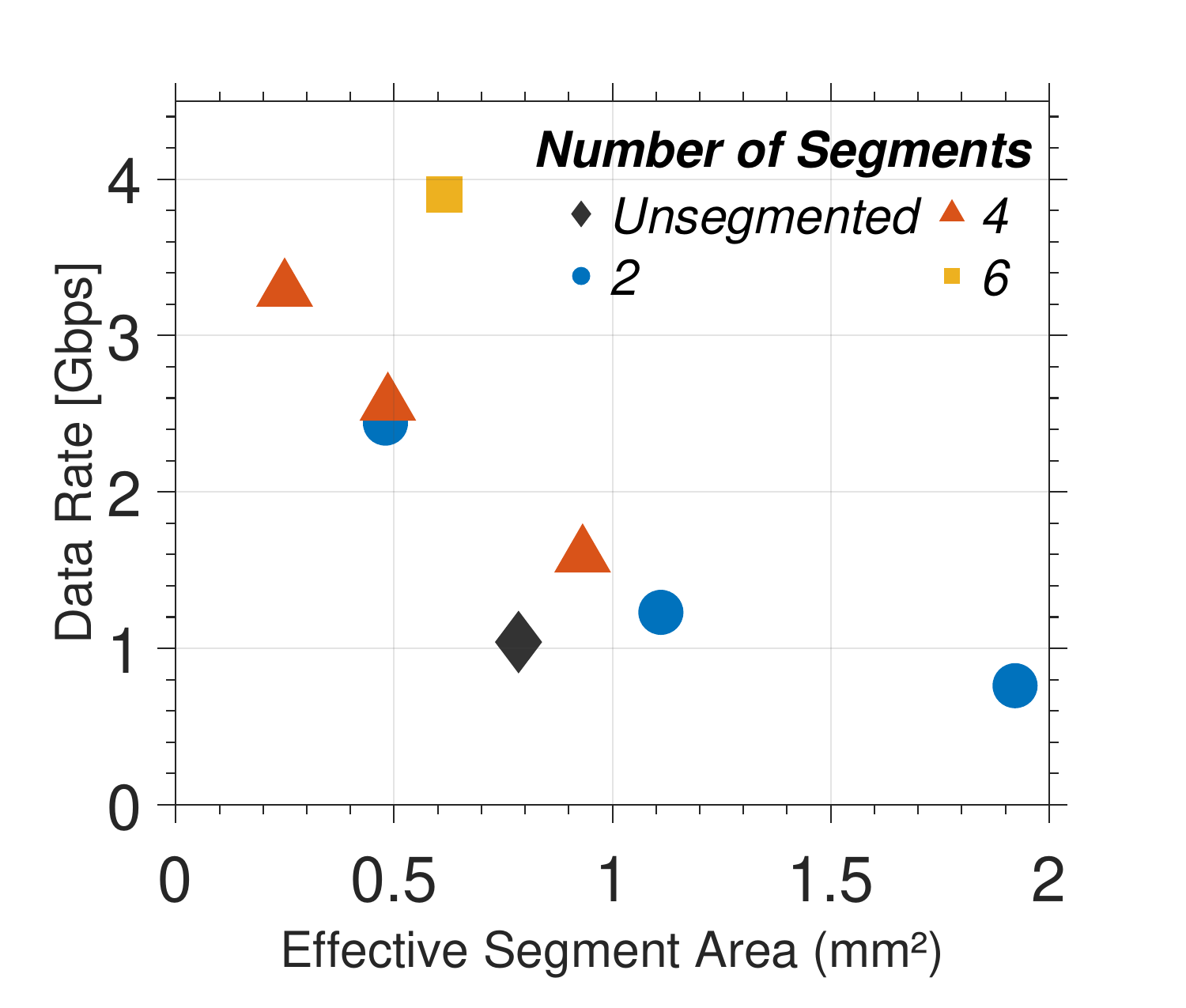}
    \caption{Data rate obtained against the effective sege ment area of the \ac{PPC}-based \ac{GaAs} receiver.}
    \label{fig:area}
\end{figure}

Figure~\ref{fig:area} presents a scatter plot of the multi-segment \ac{PPC}, showing that increasing the number of segments while reducing the effective segment area (measured in mm\textsuperscript{2}) consistently results in higher data rates, reflecting a trade-off between photodiode capacitance and the performance gains from using multiple segments concurrently. Likewise, also \ac{SNR} values generally improve with increasing segmentation with reduced signal interference (see Fig. \ref{fig:Fig_IV_Nature}).

The \ac{I-V} curves of multi-segment \ac{GaAs} \ac{PV} cells, evaluated under an optical power input of 2.3~mW (refer to \textit{Experimental setup}), describe the relationship between current and voltage, demonstrating the cells' power output and efficiency across various segment configurations. These curves illustrate how the operating condition, particularly the \ac{Pmp}, shifts with changes in segment area, number of segments, and total cell area, optimizing harvested power. Hence, the value of the resistive load is adjusted to align the device's operating condition with the \ac{Pmp}, thereby maximizing energy harvesting efficiency. The performance results are summarised in Table~\ref{tab:performance_metrics}. 
Among the two-segment devices, S(2), M(2), and L(2) achieved harvested powers of 0.49~\text{mW}, 0.89~\text{mW}, and 0.89~\text{mW}, respectively. For the four-segment configuration, S(4), M(4), and L(4) produced 0.39~\text{mW}, 0.59~\text{mW}, and 0.67~\text{mW}. The six-segment device, L(6), yielded the lowest output at 0.23~\text{mW}, likely due to non-ideal alignment and inhomogeneous illumination, as illustrated in Fig.~\ref{fig:Fig_IV_Nature}.
 
One method for analyzing the degree of current mismatch across individual segments involves analyzing the ratio of current at the maximum power point to short circuit current ($I_{\mathrm{mp}}/I_{\mathrm{sc}}$). Moreover, the \ac{PCE}, quantifying the efficiency of conversion from incident optical power to electrical output power, serves as a key comparative parameter across a range of cell areas and segment configurations. For two-segment cells, $I_{\mathrm{mp}}/I_{\mathrm{sc}}$ ratios are \SI{96.2}{\percent} (\SI{0.79}{mm^2}), \SI{99.5}{\percent} (\SI{1.77}{mm^2}), and \SI{97.2}{\percent} (\SI{3.40}{mm^2}), with corresponding \ac{PCE} values of \SI{22.1}{\percent}, \SI{38.7}{\percent}, and \SI{39.7}{\percent}. Four-segment cells show lower $I_{\mathrm{mp}}/I_{\mathrm{sc}}$ ratios of \SI{85.0}{\percent}, \SI{90.1}{\percent}, and \SI{89.5}{\percent} for the same areas, with \ac{PCE} values of \SI{19.8}{\percent}, \SI{28.3}{\percent}, and \SI{32.5}{\percent}. The six-segment configuration at \SI{3.40}{mm^2} exhibits the lowest $I_{\mathrm{mp}}/I_{\mathrm{sc}}$ ratio (\SI{66.1}{\percent}), indicating non-ideal alignment and consequently \ac{PCE} (\SI{15.1}{\percent}).  

\subsection{Analysis and Discussion}
Increasing the number of \ac{PPC} segments reduces the active area per subcell, which in turn lowers the parasitic capacitance of each subcell as well as of the series connected device, thereby enabling high-frequency operation. This is reflected in the experimentally observed communication bandwidth, which nearly doubles from 0.49~GHz to nearly 1~GHz when scaling from 2 to 6 segments in the largest cell configuration. By operating near the short-circuit condition and employing segmentation, carrier extraction is accelerated, and capacitive effects are minimized, resulting in a data rate of 3.8 Gbps—four times higher than the previous world record for similar \ac{PPC} systems. This leap in performance highlights the efficacy of segmentation and parallel load optimization in enhancing bandwidth and \ac{SNR}, positioning multi-segment \ac{GaAs} \ac{PPC} as a unique solution for high-rate data reception in compact, eye-safe \ac{SLIPT} systems. However, increased segmentation may pose new challenges for efficient operation, as discussed below, which must be carefully managed for energy-critical applications.

From an energy harvesting perspective, the $I_{\mathrm{mp}}/I_{\mathrm{sc}}$ ratio can reveal current mismatch challenges in segmented \ac{PV} cells. Current matching refers to the condition in which all series‑connected segments generate the same photocurrent, which typically requires uniform light distribution across every segment. Under these conditions, the device can operate near its theoretical maximum power. Two-segment cells exhibit higher measured $I_{\mathrm{mp}}/I_{\mathrm{sc}}$ ratios, indicating better current matching. However, increasing to four or six segments spatial illumination variations—originating from non-uniform light intensity across the cell surface caused by beam profile inhomogeneity, optical misalignment, or partial shading—resulted in decreased uniformity among the segments, and hence reduced current matching, lower $I_{\mathrm{mp}}/I_{\mathrm{sc}}$ ratio, and decreased \ac{PCE}. 
The six-segment configuration, in our experiments suffers the most significant current mismatch and lowest \ac{PCE} due to series-connected segments being limited by the least-illuminated segment. This suggests that while segmentation boosts communication performance, it complicates uniform illumination and alignment, which are critical for maintaining energy harvesting efficiency. For applications like aerospace or remote sensing, where both high data rates and power efficiency are essential, these findings underscore the need for precise optical alignment or advanced illumination compensation techniques to balance these trade-offs. A detailed comparison with state-of-the-art \ac{PV} technologies for energy-harvesting wireless communication systems is provided in Table~\ref{tab:previous_owc_pv}

\begin{table*}[!]
\centering
\small 
\caption{\scriptsize \textbf{Summary of studies on optical wireless communications using solar/\ac{PV} cells as data detectors and using \ac{OFDM} modulation.}} 
\label{tab:previous_owc_pv}
\begin{tabular}{p{2.8cm} p{1.4cm} p{1.3cm} p{1.4cm} p{1.3cm} p{1.5cm} p{1cm} p{1cm} l} 
\toprule
\textbf{Devices} & \multicolumn{5}{c}{\textbf{Wireless Communication}} & \multicolumn{2}{c}{\textbf{Energy harvested}} & \textbf{Ref.} \\
      \cmidrule(lr){1-1}     \cmidrule(lr){2-6} \cmidrule(lr){7-8} \cmidrule(lr){9-9}
\textbf{Photovoltaic technology} & \textbf{Active area (mm$^2$)} & \textbf{Comm. BW (MHz)} & \textbf{Adapt. bit \& power loading} & \textbf{Data rate (Mbps)} & \textbf{BER} & \textbf{Power (mW)} & \textbf{\ac{PCE} (\%)} & \\
\midrule
Unsegmented GaAs cell & 0.78 & 350 & Yes & 1041 & 2.2$\times$10$^{-3}$ & 1.5 & 41.7 & \cite{ref22} \\
PTB7:PC71BM organic cell & 8 & 10 & Yes & 42 & 1.1$\times$10$^{-3}$ & 0.43 & {-} & \cite{ref34} \\
Mono-crystalline \ac{Si} panel with 11 cells & 3850 & 9 & Yes & 17.05 & 1.1$\times$10$^{-3}$ & {-} & {-} & \cite{ref35} \\
Mono-crystalline \ac{Si} cell & 750 & 5 & No (16-\ac{QAM}) & 15  & 1.68$\times$10$^{-3}$ & {-} & {-} & \cite{ref36} \\
Multi-crystalline \ac{Si} panel with 36 cells & 42970 & 2 & Yes & 12 & 1.6$\times$10$^{-3}$ & 30 & {-} & \cite{ref11} \\
$n$-segment \ac{GaAs} cell & 1.7-3.5 & 490-960 & Yes & 490-3800 & 1.6$\times$10$^{-3}$ & {$\le 0.886$} & 15.1-39.7 & This work \\
\bottomrule
\end{tabular}
\end{table*}

For instance, a conventional unsegmented \ac{GaAs} cell (0.78 mm² active area) achieves a 125 MHz bandwidth, a 522 Mbps data rate, and 42\% \ac{PCE}, delivering 1.5~mW from 3.57~mW input power, ideal for compact, high-speed devices~\cite{ref1}. In contrast, an organic PTB7:PC71BM cell (8~mm$^2$) supports a 42~Mbps data rate with 0.43~mW output, suitable for low-power electronics~\cite{ref15}. Larger mono-crystalline \ac{Si} panels (e.g., 3850~mm$^2$, 26\% \ac{PCE}) achieve lower bandwidths of 9~MHz and data rates of 17~Mbps, prioritizing energy harvesting over communication speed~\cite{ref15}. These comparisons highlight that our multi-segment \ac{PPC} design excels in high-speed communication while maintaining competitive energy harvesting, though at the cost of more demanding alignment and optics or reduced \ac{PCE} with higher segmentation. This trade-off suggests that applications requiring maximum power efficiency may favor fewer segments or unsegmented designs, while high-bandwidth applications benefit from increased segmentation. Note that vertically stacked multi‑junction \ac{PPC} devices \cite{4752734, ref15, pellegrino2025simultaneous}, which combine the benefits of integrated series connection and unsegmented light reception. However, they are more sensitive to changes in wavelength and temperature, which can cause current mismatch between the subcells. These devices are noted here, but fall outside the scope of this study.
In summary, the multi-segment \ac{PPC} architecture significantly advances the state-of-the-art in infrared optical wireless communication by mitigating capacitance limitations, enabling compact, eye-safe devices with higher bandwidth and data rates as compared with unsegmented cells. The ability to independently tune optical absorption, carrier extraction, and parasitic capacitance per segment offers flexibility for tailoring performance to specific needs, such as high-speed data transfer for backhaul systems or efficient energy harvesting in remote sensing. However, the observed reduction in \ac{PCE} with increased segmentation highlights the importance of optimizing illumination uniformity and alignment to maximize both communication and energy harvesting performance in practical deployments.

\section{Conclusion}

This work demonstrates the efficacy of multi-segment \ac{GaAs}-based \acp{PPC} in \ac{SLIPT} systems for high-speed, energy-efficient optical wireless communication. By segmenting the active area into 2, 4, or 6 subcells, we achieved a record 3.8~Gbps data rate, a fourfold improvement over prior works, with up to 39.7\% power conversion efficiency from a 2.3~mW optical input. Segmentation reduces capacitance, enabling bandwidths near 1~GHz for the 2.08~mm cell with 6 segments, positioning \acp{PPC} as enablers for 6G networks.
{\appendix [Eye saftey assessment]

To ensure ocular safety in the proposed optical wireless power transfer system, this study evaluates compliance with the \ac{MPE} limits established by the IEC 60825-1:2022 standard~\cite{ref25}. The system employs a \ac{VCSEL} operating at a wavelength of \SI{850}{\nano\meter}, integrated with collimating and focusing optics. The \ac{MPE}, also known as the exposure limit, represents the maximum irradiance level to which the eye or skin can be exposed without causing immediate or long-term harm~\cite{ref25}. The determination of the \ac{MPE} primarily depends on the exposure time, the apparent angular subtense of the source, and the wavelength of operation. According to the IEC standard~\cite{ref25,5_iec60825-1_2014}, the exposure duration ranges from \SI{100}{\femto\second} to \SI{30}{\kilo\second} (\SI{8.33}{\hour}). Adopting a conservative approach, a maximum exposure duration of \SI{8.33}{\hour} is considered in this evaluation. Light source classification is determined by the angular subtense, denoted as $\alpha$. A light source is classified as a point source if its $\alpha$ satisfies $\alpha < \alpha_{\text{min}}$, where $\alpha_{\text{min}} = \SI{1.5}{\milli\radian}$. In contrast, a source is classified as an extended source when $\alpha \geq \alpha_{\text{min}}$. Extended sources are further subdivided into two categories on the basis of their $\alpha$. Sources with $\alpha_{\text{min}} < \alpha < \alpha_{\text{max}}$ are classified as intermediate sources, while those with $\alpha \geq \alpha_{\text{max}}$, where $\alpha_{\text{max}} = \SI{100}{\milli\radian}$, are classified as large sources. In addition, components such as lenses or diffusers may be utilized to transform a point source into an extended source by increasing its $\alpha$ ~\cite{5_iec60825-1_2014}. Hence, a \ac{VCSEL}-lens setup exhibits a $\alpha > \SI{1.5}{\milli\radian}$, and thus behaves as an extended source system. The $\alpha$ is calculated as:

\begin{equation}
    \alpha = 2 \tan^{-1} \left( \frac{D_s}{2Z} \right),
\end{equation}

where $D_s$ is the source size and $Z$ is the distance from the source to the observer. For retinal hazard assessment in \ac{MPE} evaluations, the minimum evaluation distance is typically set to \SI{10}{\centi\meter}~\cite{2_international}.

With respect to wavelength, the \ac{MPE} for ocular safety is categorized by IEC standards into three regions of \SI{700}{\nano\meter}–\SI{1050}{\nano\meter}, \SI{1050}{\nano\meter}–\SI{1150}{\nano\meter}, and \SI{1150}{\nano\meter}–\SI{1400}{\nano\meter}~\cite{ref2}. The operating wavelength of our system is \SI{850}{\nano\meter}, which falls within the \SI{700}{\nano\meter}–\SI{1050}{\nano\meter} category.

The beam at the detector has a diameter $D_s$ of \SI{35}{\milli\meter}, and the observation distance $Z$ is \SI{100}{\milli\meter}. Based on Equation~2, the resulting $\alpha$ is \SI{346.4}{\milli\radian}, which confirms the source classification as an extended source. The \ac{MPE} for extended sources at \SI{850}{\nano\meter} is given by:

\begin{equation}
    \text{MPE} = 18 \times C_4 \times C_6 \times t^{-0.25},
\end{equation}

where $C_4$ and $C_6$ are correction factors for wavelength and angular subtense, respectively, and $t$ is the duration of the exposure in seconds. The wavelength correction factor $C_4$ is calculated as $ 10^{0.002 (\lambda - 700)}$, where $\lambda$ is the wavelength in nanometers~\cite{2_international}. In addition, $C_6$ is defined as $\frac{\alpha_\text{max}}{\alpha_\text{min}}$ for $\alpha > \alpha_\text{max}$. Therefore, the computed \ac{MPE} is  \SI{181.84}{\watt\per\meter\squared}. To verify compliance, the received irradiance $E$ at the human pupil is compared to \ac{MPE}, and is given by:

\begin{equation}
    E = \frac{P_r}{\pi r_p^2},
\end{equation}

Where $P_r$  is the optical power received, was measured as \SI{80}{\micro\watt} using a Thorlabs S121C power sensor, which was fitted with a \SI{7}{\milli\meter} aperture that match the typical diameter of the human pupil, and is positioned at a distance of \SI{10}{\centi\meter}. The dilated eye pupil radius $r_p$ is assumed to be \SI{3.5}{\milli\meter}, which corresponds to an irradiance $E$ of \SI{2.08}{\watt\per\meter\squared}. This value is significantly lower than the calculated \ac{MPE} of \SI{181.84}{\watt\per\meter\squared}, providing a safety margin of approximately $181.84 / 2.08 \approx 87.42$ times below the permissible limit. Therefore, the system operates well within ocular safety limits under the specified conditions. Note that the regions between the \ac{VCSEL} and the collimation lens, as well as between the focusing lens and the \ac{PV} cell, require enclosures to prevent potential exposure to higher irradiance in these uncontained areas, in accordance with IEC 60825-1.}

\section*{Acknowledgements}
This work is supported in part by the Green Optical Wireless Communications Facilitated by Photonic Power Harvesting (GreenCom) [EP/X027511/1], the Fraunhofer ICON grant, and the Platform for Driving Ultimate Connectivity (TITAN) extension, [EP/Y037243/1]. The authors thank ISE colleagues David Lackner for epitaxial growth, Ranka Koch and Eduard Oliva for clean room wafer fabrication, Rok Kimovec for taking measurements, and Gerald Siefer and Andreas Bett for the valuable discussions.



\vspace{0.3cm}

\section*{List of Acronyms}
{\renewcommand{\itemsep}{\parsep}
\begin{acronym}[UML]
    \acro{AC}[AC]{alternating current}
    \acro{ARC}[ARC]{anti-reflection coating}
    \acro{AWG}[AWG]{arbitrary waveform generator}
    \acro{BER}[BER]{bit error rate}
    \acro{CdTe}[CdTe]{cadmium telluride}
    \acro{DC}[DC]{direct current}
    \acro{DCO}[DCO]{DC-biased optical}
    \acro{EH}[EH]{energy harvesting}
    \acro{EM}[EM]{electromagnetic}
    \acro{FEC}[FEC]{forward error correction}
    \acro{FFT}[FFT]{Fast Fourier Transform}
    \acro{FSO}[FSO]{free space optics}
    \acro{GaAs}[GaAs]{gallium arsenide}
    \acro{GaInP}[GaInP]{gallium indium phosphide}
    \acro{IFFT}[IFFT]{Inverse fast Fourier transform}
    \acro{IoT}[IoT]{Internet of Things}
    \acro{LED}[LED]{light-emitting diode}
    \acro{MIM}[MIM]{monolithic interconnected module}
    \acro{MIMO}[MIMO]{multiple-input multiple-output}
    \acro{MOVPE}[MOVPE]{metal-organic vapor phase epitaxy}
    \acro{MPE}[MPE]{maximum permissible exposure}
    \acro{OFDM}[OFDM]{orthogonal frequency-division multiplexing}
    \acro{OWC}[OWC]{optical wireless communication}
    \acro{PC}[PC]{personal computer}
    \acro{PCE}[PCE]{power conversion efficiency}
    \acro{Pmp}[$P_{\text{mp}}$]{power at maximum power point}
    \acro{PPC}[PPC]{photonic power converter}
    \acro{PV}[PV]{photovoltaic}
    \acro{QAM}[QAM]{quadrature amplitude modulation}
    \acro{RF}[RF]{radio frequency}
    \acro{Rx}[Rx]{receiver}
    \acro{Si}[Si]{silicon}
    \acro{SLIPT}[SLIPT]{simultaneous lightwave information and power transfer}
    \acro{SNR}[SNR]{signal-to-noise ratio}
    \acro{SWIPT}[SWIPT]{simultaneous wireless information and power transfer}
    \acro{VCSEL}[VCSEL]{vertical-cavity surface-emitting laser}
    \acro{I-V}[I-V]{current-voltage}
\end{acronym}}

\bibliographystyle{IEEEtran}

\bibliography{references}  

\label{LastPage}
\end{document}